\documentclass{appolb}
\usepackage{bbm}
\usepackage{graphicx}
\usepackage{epsfig}
\usepackage{bbm}
\usepackage{amsfonts}
\usepackage{psfrag}

\newcommand{\beq}{\begin{equation}}
\newcommand{\eeq}{\end{equation}}
\newcommand{\beqa}{\begin{eqnarray}}
\newcommand{\eeqa}{\end{eqnarray}}
\newcommand{\abs}[1]{\left\vert#1\right\vert}
\renewcommand{\b}{\circ}
\renewcommand{\d}{{\rm d}}
\newcommand{\ds}{\displaystyle}
\renewcommand{\e}{{\rm e}}
\newcommand{\eq}{{\rm eq}}
\newcommand{\frad}[2]{\ds{\frac{#1}{#2}}}
\renewcommand{\i}{{\rm i}}
\renewcommand{\max}{{\rm max}}

\renewcommand{\min}{{\rm min}}
\newcommand{\n}{\bullet}
\newcommand{\tr}{\mathop{\rm tr}\nolimits}
\newcommand{\w}{\widehat}
\newcommand{\A}{{\rm A}}
\newcommand{\B}{{\rm B}}
\newcommand{\C}{{\rm C}}
\newcommand{\D}{{\rm D}}
\newcommand{\Frac}{\mathop{\rm Frac}\nolimits}
\newcommand{\GRW}{{\rm GRW}}
\newcommand{\Int}{\mathop{\rm Int}\nolimits}
\newcommand{\MERW}{{\rm MERW}}
\newcommand{\RW}{{\rm RW}}

\begin{document}
\date{\today}
\eqsec
\pagestyle{plain}
\newcount\eLiNe\eLiNe=\inputlineno\advance\eLiNe by -1

\title{The various facets of random walk entropy\footnote{Based on a lecture presented by Z.B. at the 22nd Marian Smoluchowski Symposium on Statistical Physics (Zakopane, Poland, September 12--17, 2009).}}

\author{Z. Burda, J. Duda\address{Marian Smoluchowski Institute of Physics,
Jagellonian
University, Reymonta 4, 30-059 Krak\'ow, Poland}
\and J.M. Luck\address{Institut de Physique Th\'eorique, CEA IPhT and CNRS URA
2306,
CEA Saclay, 91191 Gif-sur-Yvette cedex, France}
\and B. Waclaw\address{SUPA, School of Physics and Astronomy, University of
Edinburgh, Mayfield Road, Edinburgh EH9 3JZ, UK}
}
\maketitle

\begin{abstract}
We review various features of the statistics of random paths on graphs.
The relationship between path statistics and Quantum Mechanics (QM)
leads to two canonical ways of defining random walk on a graph,
which have different statistics and hence different entropies.
Generic random walk (GRW) is in correspondence with the field-theoretical formalism,
whereas maximal entropy random walk (MERW), introduced by us in a recent work,
is motivated by the Feynman path-integral formulation of QM.
GRW maximizes entropy locally (neighbors are chosen with equal probabilities),
in contrast to MERW which does so globally
(all paths of given length and endpoints are equally probable).
The stationary distribution for MERW
is given by the ground state of a quantum-mechanical problem
where nodes whose degree is smaller than average act as repulsive impurities.
We investigate static and dynamical properties GRW and MERW in a variety of examples
in one and two dimensions.
The most spectacular difference arises in the case of weakly diluted lattices,
where a particle performing MERW gets eventually trapped
in the largest nearly spherical region which is free of impurities.
We put forward a quantitative explanation of this localization effect
in terms of a classical Lifshitz phenomenon.
\end{abstract}


\section{Introduction}
\label{sec1}

This paper presents a review of various facets of
the statistics of random walks on graphs and other geometrical structures.
Brownian motion has been proposed in the seminal works by
Einstein~\cite{ref:ein}
and Smoluchowski~\cite{ref:smol} as a microscopic theory for diffusive
transport.
Random walk (RW) models stem from a discretization of Brownian motion.
The discretization procedure can be motivated either on theoretical grounds
(it provides a cutoff regularizing the short-distance singularities
which plague the continuous Brownian motion)
or by practical considerations (it makes numerical simulations easier).
There are many incarnations of RW,
where either space or time is separately considered as discrete or continuous.
The most celebrated ones include the Polya walk on a lattice
and continuous-time random walk (CTRW)~\cite{ref:hughes}.
RW is such a natural construction
that it has altogether been used across the whole realm of sciences.
We refer the reader to~\cite{ref:feller} for an exposition of the probabilistic
foundations
of Brownian motion and RW, and to~\cite{ref:mw} for a discussion of a range
of applications in the physical sciences.

In this paper we consider for definiteness discrete-time RW on
lattices (or graphs in general).
Within this context, RW is a Markov chain which
describes the stochastic trajectory of a particle (random walker) taking
successive random steps.
For instance, in the well-known case of the Polya walk on a lattice,
at each time step the particle jumps at random onto one of the neighboring nodes.
The walk thus generates a random path on the lattice.

The relationship between RW and the statistics and entropy of paths
is the main thread of this paper.
In Section~\ref{sec2} we review various features of the statistics of paths.
We recall how to count paths by means of the adjacency matrix $A$ of a graph.
We then present the relationship between RW and path integrals.
The Feynman path-integral formalism~\cite{ref:feynman} for a free particle,
where trajectories are weighted only by their length,
suggests that all paths of given length and endpoints should be equally
probable.
At variance with this picture,
the field-theoretical approach to a relativistic particle
propagating in a curved background requires that paths of the same length are not equally probable,
but rather that their statistical weights depends on the nodes through which
they pass.
Both formalisms only agree in the particular case of $k$-regular graphs,
where all the nodes have the same degree.
Section~\ref{sec3} is devoted to RW models on an arbitrary graph.
The main emphasis is put on two different canonical ways of defining RW,
namely generic random walk (GRW) and maximal entropy random walk (MERW),
respectively corresponding to the field-theoretical
and path-integral formalisms.
In generic random walk (GRW),
the particle sitting at a node of degree $k$ jumps onto any neighboring node
with uniform probability $1/k$, maximizing thus the entropy production locally,
albeit not globally.
In the stationary state, the probability $\pi_a$ of finding the particle at
node $a$
is proportional to the degree $k_a$.
When the lattice is regular (\ie, all nodes have the same degree),
all paths of a given length between two given points are equally probable,
and thus have maximal entropy.
Nevertheless, in accord with the field-theoretical formalism,
as soon as the graph is not regular GRW trajectories are no longer
equally probable.

We then turn to maximal entropy random walk (MERW).
This novel kind of RW has been put forward and investigated by us in a recent
work~\cite{ref:us}.
It is defined in such a way as to ensure that all paths of given length and
endpoints are equally probable, in accord with the Feynman path integral.
In other words, MERW is meant to maximize entropy globally, albeit not locally.
It is still a Markov (memoryless) process,
defined by local but non-trivial rules involving the largest eigenvalue
of the adjacency matrix of the graph
and the associated Perron-Frobenius eigenvector.
The latter appears as the ground state of a quantum-mechanical
tight-binding Hamiltonian,
where the nodes $a$ whose degree $k_a$ is smaller than the maximal degree
$k_\max$
carry a repulsive site potential $V_a=k_\max-k_a$.
In the stationary state, the probability $\pi_a$ of finding the particle at
node $a$
is given by the square of the component $\psi_{1a}$ of the Perron-Frobenius
eigenvector.
As a byproduct, we introduce three definitions of the effective degree of a
graph.
The connection between RW, and especially GRW and MERW,
and stochastic quantization is then made in Section~\ref{sec4}.

In the rest of the paper we make a comparative investigation of GRW and MERW.
A variety of examples of finite graphs are dealt with in Section~\ref{sec5},
including bipartite and linear graphs.
Section~\ref{sec6} is devoted to extended one-dimensional structures (ladder
graphs).
As a general rule, the effect of the repulsive potential
on a particle performing MERW extends over the whole system.
We consider successively the situation of one or two repulsive impurities,
the converse situation of an attractive impurity where the stationary
distribution
is carried by a localized impurity state,
and the case of a weakly diluted graph,
obtained by removing a small fraction of bonds at random.
The higher-dimensional situation
is illustrated by several two-dimensional examples in Section~\ref{sec7}.
It is shown that MERW can lead to better transport properties than GRW,
on the example of a non-bipartite two-dimensional lattice, the dual $(4,8^2)$
lattice.
In the case of weakly diluted lattices,
obtained by removing a small fraction of bonds at random,
it is shown that any small amount of disorder is sufficient to localize
the stationary state of MERW in the largest nearly spherical region
which is free of defects.
This unexpected localization effect takes place in any dimension.
We provide a quantitative explanation for it
in terms of a classical Lifshitz phenomenon.

The interested reader is referred to the interactive MATHEMATICA
demonstration by one of us (BW) for many more illustrations of the unusual
static and dynamical features of MERW~\cite{ref:math}.

\section{Statistics of paths}
\label{sec2}

\subsection{Enumerating paths on a graph}

A path is a very common object in graph theory
(see, \eg,~\cite{ref:wilson},~\cite{ref:bollobas}).
It is a finite sequence of adjacent (neighboring) nodes on a graph.
Its length is defined as the number $n$ of steps of the path.
Each step follows a link (bond, edge) of the graph.

In this paper we consider only undirected graphs,
\ie, graphs whose edges have no orientation.
We first recall how to enumerate paths $\{\gamma_{ba}(n)\}$
of length $n$ going from node $a$ to node $b$ on a finite connected graph.
Throughout this paper we use notations consistent with those used
in QM: the initial (resp. final) state is the second (resp.
first)
index of transition matrices or propagators.
The number $N_{ba}(n)$ of paths
$\{\gamma_{ba}(n)\}$ can be calculated recursively as
\begin{equation}
N_{ba}(n+1) = \sum_c A_{bc} N_{ca}(n),
\label{recurrence}
\end{equation}
where $A=(A_{ab})$ is the adjacency matrix of the graph:
\begin{equation}
A_{ab} = \left\{ \begin{array}{ll} 1 & {\rm if} \ a,b \ \hbox{are neighbors,}\\
0 & \hbox{otherwise.} \end{array} \right.
\end{equation}
For a finite undirected graph with $N$ nodes, $A$ is a symmetric $N\times N$
matrix.
We have
\beq
\sum_bA_{ab}=k_a,
\eeq
where $k_a$ denotes the degree (number of neighbors) of node $a$, and
\beq
\tr A^2=\sum_ak_a=2L,
\label{traces}
\eeq
where $L$ is the number of links of the graph.

Applying the recursion relation (\ref{recurrence}) $n$ times
to the initial condition $N_{ba}(0) = \delta_{ba}$,
where $\delta_{ba}$ is the Kronecker delta, one obtains
\begin{equation}
N_{ba}(n) = (A^n)_{ba} = \sum_{i} \psi_{ib}\psi_{ia} \lambda_i^n,
\label{Nba}
\end{equation}
where $A^n$ is $n$-th power of the adjacency matrix,
and $\psi_i$ denote the normalized eigenvectors of $A$ associated
with the eigenvalues $\lambda_i$:
\begin{equation}
A\psi_i = \lambda_i \psi_i,\qquad \sum_a\psi_{ia}^2=1.
\label{Apsi}
\end{equation}
The second index of $\psi_{ia}$ denotes $a$-th component of $\psi_i$.
The $N$ eigenvalues are assumed to be ordered so as to have decreasing absolute
values:
$|\lambda_1| \ge |\lambda_2| \ge \ldots$
In the large-$n$ limit, the sum
(\ref{Nba}) is dominated by the largest eigenvalue $\lambda_1$ of the
adjacency matrix.

After Boltzmann, Gibbs, and Shannon, the entropy $H_{ba}(n)$ of the ensemble of paths
$\{\gamma_{ba}(n)\}$ is defined as the logarithm of the number of paths:
\beq
H_{ba}(n)=\ln N_{ba}(n).
\eeq
For $n \rightarrow \infty$, this entropy grows as
\begin{equation}
H_{ba}(n) \approx n \ln \lambda_1+\ln(\psi_{1b}\psi_{1a}).
\label{hba}
\end{equation}
The leading term is independent of the positions of the endpoints $a,b$.
So, for large $n$, the entropy density per step (also called entropy production
rate),
\begin{equation}
h =\lim_{n\to\infty}\frac{H_{ba}(n)}{n}=\ln\lambda_1,
\label{slambda}
\end{equation}
only depends on the largest eigenvalue $\lambda_1$ of the adjacency matrix $A$.
The concept of entropy production can be generalized to a broader
class of Markov chains, where the transition probabilities
also depend on some field defined on the graph~\cite{ref:gomez}.

Under the mild hypothesis that the adjacency matrix $A$ is {\it primitive} (or {\it regular}, see
\cite{ref:bellman}),
it follows from the Perron-Frobenius theorem that $\lambda_1$ is positive,
whereas all the other eigenvalues are strictly smaller in modulus,
and that the corresponding eigenvector $\psi_1$ can be chosen so as to
have strictly positive components: $\psi_{1a}>0$.
In the present situation of undirected graphs,
the matrix $A$ is symmetric and its spectrum is real.
The primitiveness hypothesis thus only excludes the case of bipartite graphs,
which can be dealt with separately (see Section~\ref{sec:bip}).
For a bipartite graph,
we have $\lambda_2=-\lambda_1$, and hence $N_{ba}(n)$ oscillates with $n$,
so that Eq.~(\ref{hba}) possesses an additive oscillating contribution of order unity.

\subsection{\label{frp}The path integral of a free relativistic particle}

An interesting application of the statistics of paths
is the description of a free relativistic particle.
In the Feynman formulation of QM,
quantum amplitudes are calculated as path integrals~\cite{ref:feynman}.
For a particle propagating in
$d$-dimensional Minkowski spacetime, the amplitude is given as an integral
over spacetime trajectories $x_{ba}(\tau)$ going from an initial spacetime point $a$
to a final one $b$:
\begin{equation}
G_{ba} = \int [\D x_{ba}(\tau)]\ \e^{ \frac{\i}{\hbar} S[x_{ba}(\tau)]},
\label{gba}
\end{equation}
where $\hbar$ is Planck's constant. The action of a free scalar particle
is $S= M c\int_a^b \d s$, where ${\rm d} s^2 = ({\rm d}x^0)^2 - ({\rm
d}\vec{x})^2$ is the spacetime interval, and $M$ is the mass of the particle.
The spacetime coordinates of point $x=(ct,\vec{x})$ will be denoted by $x^\mu$
($\mu=0,1,\ldots,d-1$) and the speed of light as well as Planck's
constant will be set to unity for convenience: $\hbar=c=1$. One way of
performing
the integral (\ref{gba}) is to use the Wick rotation $(x^0,\vec{x}) \rightarrow
(\i x^0,\vec{x})$ and to calculate a related quantity, called the Euclidean
propagator
(or Euclidean kernel).
One then rotates the result back to the Minkowskian sector. Under Wick's
rotation,
the spacetime interval ${\rm d}s^2= ({\rm d}x^0)^2 - ({\rm d}\vec{x})^2$
transforms to $-{\rm d}s^2 = -\left(({\rm d}x^0)^2 + ({\rm
d}\vec{x})^2\right)$. The Euclidean propagator is defined by
taking a proper branch of the square root of $\sqrt{-{\rm d} s^2}$:
\begin{equation}
G_{ba} = \int [\D x_{ba}(\tau)]\ \e^{-S[x_{ba}(\tau)]},
\label{GEba}
\end{equation}
where the free-particle action
\begin{equation}
S[x(\tau)] = M \int\d\tau \sqrt{\eta_{\mu\nu} \dot{x}^\mu\dot{x}^\nu} = M
\int\d s
\end{equation}
is proportional to the length of the corresponding Euclidean trajectory.
Here
$\eta = (\eta_{\mu\nu}) = {\rm diag}(1,\ldots,1)$ denotes the Euclidean metric
tensor and the Einstein summation convention is used. The trajectory
$\tau\mapsto x(\tau)$
is parametrized by its proper time $\tau$. The dots denote
derivatives with
respect to $\tau$. The action $S[x(\tau)]$
is invariant with respect to diffeomorphic reparameterizations of the
trajectory,
so that the choice of parametrization does not play any role.

\begin{figure}[!ht]
\psfrag{x}{$x$}\psfrag{t}{$t$}
\begin{center}
\includegraphics[width=10cm]{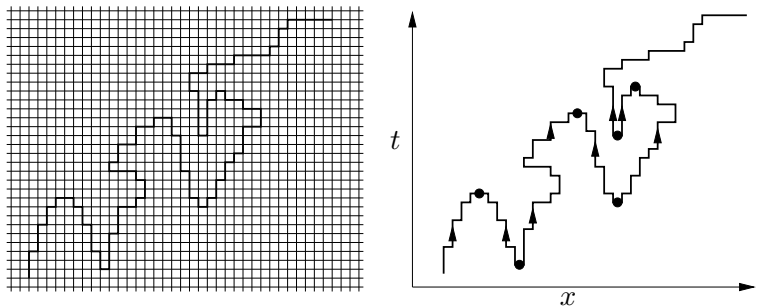}
\end{center}
\caption{\label{fig1}
Left: Example of a path on a 2D lattice. Right: the same
path is interpreted as a trajectory of a relativistic quantum particle in
(1+1)-dimensional
spacetime. Arrows show the direction of time, whereas black circles are
creation/annihilation events.}
\end{figure}

The Euclidean propagator (\ref{GEba}) calls for a probabilistic interpretation,
as the integrand is positive. A good way of emphasizing this feature
is to use a lattice regularization. This is a standard strategy in field theory: one
discretizes spacetime, calculates the propagator, and then takes the
continuum limit by sending the lattice spacing to zero (see, \eg,~\cite{ref:id}).
If one does this carefully, the outcome is independent of the discretization and
all the symmetries of the underlying continuum theory are restored.
Finally, the results are rotated
from the Euclidean sector back to the Minkowskian one. In order to show how this
works,
let us consider the case where the continuous spacetime
$\mathbb{R}^d$ is discretized into the hypercubic lattice $\mathbb{Z}^d$,
with lattice spacing $\epsilon$.
The integral over all possible trajectories from $a$ to $b$ in the Euclidean
propagator (\ref{GEba}) is replaced by a sum over all possible paths between
$a$ and $b$:
\begin{equation}
G_{ba} = \sum_{\{ \gamma_{ba} \}} W(\gamma_{ba})\ \e^{-\mu n[\gamma_{ba}]},
\label{GFPI}
\end{equation}
where $\mu = M\epsilon$, $n[\gamma_{ba}]$ is the length of $\gamma_{ba}$, equal
to the number
of edges along this trajectory, and $W(\gamma_{ba})$ are statistical weights
corresponding to the integration measure $[\D x_{ba}(t)]$ defining the ensemble
of trajectories in Eq.~(\ref{GEba}). At variance with
the non-relativistic case, in relativistic QM
trajectories may go back and forth in the time direction. A trajectory which
locally goes
backward in time is interpreted as an antiparticle propagating forward in time.
Turning points, where the trajectory changes its time direction, correspond to
particle-antiparticle creation or annihilation events (see Figure~\ref{fig1}).

Now, an interesting question arises: how should the weights
$\{W(\gamma_{ba})\}$
be chosen in order to obtain the {\it correct} relativistic Quantum Mechanics?
The most natural choice consists in setting all $W(\gamma_{ba})$ to be equal.
In
other words, all the possible trajectories with a given length are assumed
to be equally probable.
Although this is the right prescription for paths on regular lattices,
we shall see that this choice stands in contradiction with another
formulation of
QM if the underlying lattice is not regular. Yet, let us first
consider the case where $W(\gamma_{ba})=1$. Eq.~(\ref{GFPI}) can then
be viewed as the Laplace transform
of the number of trajectories of length $n$,
\beqa
G_{ba}
&=&\sum_{n} \e^{-\mu n} N_{ba}(n) = \sum_{n} \e^{-\mu n}
\left(A^n\right)_{ba}\nonumber\\
&=&\left( \mathbbm{1} - \e^{-\mu} A\right)^{-1}_{ba}
=\sum_{i}\frac{\psi_{ib}\psi_{ia}}{1-\e^{-\mu}\lambda_i},
\label{laplace}
\eeqa
where we denoted the identity matrix by $\mathbbm{1} = (\delta_{ab})$.
The Euclidean propagator is thus directly related to the statistics of paths,
and to the adjacency matrix~{$A$}.
It has a first singularity (simple pole) for $\mu \rightarrow \ln \lambda_1$.

The generalization of the above construction to a curved spacetime is
straightforward.
In discrete quantum gravity models curved backgrounds
are discretized using simplicial manifolds~\cite{ref:dqg}.
Such a manifold consists of equilateral $d$-simplices which are put
together to form a $d$-dimensional manifold. In $d=2$ dimensions
we obtain an equilateral triangulation~\cite{ref:2qg}. One can additionally impose
a causal structure by introducing a foliation which singles out the temporal
direction.
This approach involving causal dynamical triangulations,
referred to as Lorentzian simplicial gravity~\cite{ref:lg},
is also extensively used to discretize path integrals in quantum gravity.
The Euclidean action of a free particle in a curved background is
\begin{equation}
S[x(\tau)] = M \int\d\tau \sqrt{g_{\mu\nu} \dot{x}^\mu\dot{x}^\nu} = M \int\d
s,
\end{equation}
where $g_{\mu\nu}$ is the metric tensor, so that the action is, as
before, proportional to the length of trajectory. Assuming again weights
$W(\gamma_{ba})=1$, the Euclidean propagator is given by the same formula
(\ref{laplace})
as in flat spacetime, except that now $A$ is the adjacency matrix of
the graph representing the non-trivial geometrical background
in which the particle propagates. One can alternatively consider
paths on a graph dual to the given simplicial manifold. This graph is
$k$-regular with $k=d+1$. The corresponding paths pass through
the centers of the simplices of the manifold.

\subsection{Field-theoretical approach}

Let us now discuss an alternative derivation of the Euclidean propagator.
A free relativistic particle propagating in a given background
can be described by a free field (see, \eg,~\cite{ref:zinn}) with action
\begin{equation}
S = \frac{1}{2} \int {\rm d}^d x \sqrt{g}
\left(g^{\mu\nu} \frac{\partial \Phi}{\partial x^\mu}
\frac{\partial \Phi}{\partial x^\nu} + M^2 \Phi^2 \right),
\end{equation}
where $\Phi=\Phi(x)$ is a scalar field and $g_{\mu\nu}=g_{\mu\nu}(x)$
is the metric tensor at the spacetime point $x$. We consider here only the
Euclidean sector, \ie, $g_{\mu\nu}$ having the Euclidean signature
$(+,+,\dots,+)$. The matrix $g^{\mu\nu}$ is the inverse
of $g_{\mu\nu}$, and $g = \det (g_{\mu\nu})$. Integrating the first term by
parts and assuming that boundary terms vanish, one obtains
\begin{equation}
S = \frac{1}{2} \int {\rm d}^d x \sqrt{g} \
\Phi \left( - \nabla^2 + M^2\right) \Phi,
\end{equation}
where
\begin{equation}
\nabla^2 = \frac{1}{\sqrt{g}}
\frac{\partial}{\partial x^\mu} \sqrt{g} g^{\mu\nu}
\frac{\partial}{\partial x^\nu}
\end{equation}
is the Laplace-Beltrami operator. The Euclidean propagator is defined as the two-point correlation function
\begin{equation}
G(x,y)=\frac{1}{Z} \int \left[\D \Phi\right] \Phi(x) \Phi(y)\ \e^{-S},
\end{equation}
where $Z=\int \left[\D \Phi\right] \e^{-S}$ and $\left[\D \Phi\right]$ is
the integration measure for a scalar field in the given geometrical background.
The meaning of the latter measure becomes clear if one discretizes the
geometry, using an equilateral graph with fields $\{ \phi_a \}$ located at
the nodes. The integration measure is assumed to be a product measure
$[\D \Phi] = \prod_a {\rm d} \phi_a$.
This is the same kind of assumption as was made about the weights $\{W(\gamma_{ba})\}$ in Section~\ref{frp}.
This assumption is validated by the fact that it leads to the same propagator
as the one obtained from the Klein-Gordon equation of motion of the scalar field.
The discretized action reads
\begin{equation}
S = \frac{1}{2} \sum_{ab} A_{ab}(\phi_a-\phi_b)^2 + \frac{m^2}{2} \sum_a k_a
\phi_a^2,
\end{equation}
or, equivalently,
\begin{equation}
S = \frac{1}{2} \sum_{ab} \phi_{a} \left( -\Delta_{ab} + m^2 k_a \delta_{ab}
\right)\phi_b,
\end{equation}
where $k_a$ denotes the degree of node $a$, and
\beq
\Delta_{ab} = - k_a \delta_{ab} + A_{ab}
\eeq
is the topological (or graph) Laplacian. Note that
the sign of this operator is opposite to that commonly used in graph theory.
We use this sign convention because we want
the graph Laplacian to become the Laplace operator in the continuum limit.
As a consequence, all eigenvalues of this operator are non-positive. The
correspondence between the dimensionless discretized quantities and those in
the original continuum theory is $\Phi(x) \leftrightarrow \epsilon^{1-d/2}
\phi_a$, $M^2
\leftrightarrow \epsilon^{-2} m^2$,
${\rm d}^d x \sqrt{g} \leftrightarrow \epsilon^d k_a$, where $\epsilon$ is the
lattice spacing. The Laplace-Beltrami operator is discretized
as the lattice Laplacian: ${\rm d}^d x \sqrt{g} \nabla^2 \leftrightarrow
\epsilon^{d-2} \Delta_{ab}$.

The discretized Euclidean propagator is given by the Gaussian integral
\begin{equation}
G_{ba}=\frac{1}{Z}\int\prod_c\d\phi_c\ \phi_b\phi_a\,\exp
\left(-\frac{1}{2}\sum_{cd}\phi_c\left(-\Delta_{cd}+m^2k_c\delta_{cd}\right)\phi_d\right),
\end{equation}
which can be done explicitly:
\begin{equation}
G_{ba} = \frac{1}{\sqrt{k_bk_a}}
\left( m^2 \mathbbm{1} - \widehat{\Delta} \right)^{-1}_{ba} =
\frac{\e^{-\hat{\mu}}}{\sqrt{k_bk_a}} \widehat{G}_{ba},
\end{equation}
where
\begin{equation}
\widehat{G}_{ba} =
\left(\mathbbm{1} - \e^{-\hat{\mu}} \widehat{A} \right)^{-1}_{ba}.
\label{Gba}
\end{equation}
The normalized graph Laplacian $\widehat{\Delta}$
is defined by $\widehat{\Delta}_{ba} = \Delta_{ba}/\sqrt{k_bk_a}$,
whereas $\widehat{A}$ is a matrix
with entries $\widehat{A}_{ba} = A_{ba}/\sqrt{k_bk_a}$,
and $\hat{\mu} = \ln(1+m^2)$. The geometric series expansion
of (\ref{Gba}) yields powers of the matrix $\widehat{A}$ of the
type $(\widehat{A}^n)_{ab}$, which are nonzero only if all the factors $A_{cd}$ in
the product represent adjacent edges of the graph. In other words,
$(\widehat{A}^n)_{ba}$ generates paths on the graph, each with some
statistical weight. We thus obtain
\begin{equation}
\widehat{G}_{ba} = \sum_{n} \e^{-\hat{\mu} n}
({\widehat{A}}^n)_{ba} =
\sqrt{k_bk_a} \sum_{\{\gamma_{ba}\}}
\e^{-\hat{\mu} n[\gamma_{ba}]} \prod_{c \in \gamma_{ba}}\frac{1}{k_c}.
\label{GFT}
\end{equation}

Comparing Eqs.~(\ref{laplace}) and (\ref{GFT}), we see that both approaches
lead to different propagators for a free particle on a graph
representing a discretized curved background. The main
difference is the following.
In Eq.~(\ref{laplace}) all paths have equal weights $W(\gamma_{ba})=1$,
as they are just weighted by an exponential factor depending on the path's length,
whereas in Eq.~(\ref{GFT}) paths have an additional weight
$W(\gamma_{ba})=\sqrt{k_bk_a}\prod_{c \in \gamma_{ba}} (1/k_c)$.
Thus, the path-integral formulation which is consistent with the
field-theoretical
approach requires that paths of the same length are not equally probable,
but rather that their statistical weights depends on the nodes through which
they pass.
In the particular case of $k$-regular graphs, \ie, graphs
for which all nodes have the same degree $k$,
(\ref{laplace}) and~(\ref{GFT}) are equivalent.
Indeed, since the weights
$W(\gamma_{ba})=k^{-n[\gamma_{ba}]}$ only depend on the length of paths,
the propagators can be mapped onto one another through $\mu = \hat{\mu} + \ln k$.

\section{Random walks on a graph}
\label{sec3}

So far we have discussed the statistics of paths and its relationship to path
integrals. Another area where the statistics of paths naturally
arises is random walk on graphs. Random walk is
a stochastic process, providing a microscopic representation of diffusion,
which describes a particle (or a gas of non-interacting particles)
hopping between the nodes of a graph. One is interested in the probability
$p_{ba}(\tau)$ that a particle, which was initially at node $a$,
is at node $b$ at the later time $\tau$. We will consider only
discrete-time random walks, such that the particle hops to a neighboring
node of the graph at each time step. Graphs can be treated as a discretization of
a geometrical background in which diffusion takes place. In this
case one usually wishes to restore the continuum theory by sending the lattice
spacing and the time interval to zero in a proper way ({\it diffusive scaling}).
Graphs may also model real discrete structures.
In this case there is no reason to invoke a continuum limit.
For instance, complex networks are commonly used to model the Internet,
the worldwide airline network, social networks, and so on~\cite{ref:nets}.
In this context, random walk may describe the propagation of information,
passengers, etc.

\subsection{Generalities}

Discrete-time random walk on a finite connected graph is an irreducible Markov
chain.
The stochastic motion of the particle is encoded in transition probabilities
$\{P_{ba}\}$ that a particle sitting at node $a$ will hop to node $b$ at the next time step.
One can collect the transition probabilities in a transition (or Markov)
matrix $P=(P_{ba})$ (see~\cite{ref:mm}).
The matrix $P$ fulfills the conditions for being a stochastic matrix:
the probabilities are non-negative ($P_{ba}\ge 0$),
while the entries in each
row sum up to unity: $\sum_b P_{ba}=1$, ensuring the conservation of probability.
We will only consider transition matrices which conform with the structure
of the undirected graph, that is for any pair $(a,b)$ of neighboring nodes
we have $A_{ab}=A_{ba}=1$, and $P_{ab}>0$, $P_{ba}>0$,
although $P_{ab}\ne P_{ba}$ in general.
All the other entries of $P$ (including the diagonal ones) are zero,
so that in a single time step the particle may only hop to a neighboring node.

Using the transition matrix $P$, one can write a recursive relation for
the probabilities $p_{ba}(\tau)$:
\begin{equation}
p_{ba}(\tau+1) = \sum_c P_{bc} p_{ca}(\tau),
\label{master}
\end{equation}
which is analogous to the combinatorial formula (\ref{recurrence}).
Solving it with the initial condition $p_{ba}(0)=\delta_{ba}$,
corresponding to the particle starting at node~$a$, one obtains
\begin{equation}
p_{ba}(\tau) = (P^\tau)_{ba},
\label{pba}
\end{equation}
which is again analogous to Eq.~(\ref{Nba}).
The interpretation of both formulas is different:
(\ref{Nba}) has a combinatorial meaning,
while (\ref{pba}) has a probabilistic one.
There is, however, a strong similarity between both problems,
namely that they can be formulated in terms of
path statistics. A particle performing a random walk on a graph marks a
trajectory of consecutive nodes visited during the walk. The length $n$ of this
path is equal to the time (number of steps) $\tau$ of the random walk.
Let $\gamma_{a_\tau a_0}=(a_\tau,\ldots,a_1,a_0)$ be a path of length $\tau$
from $a_0$ to $a_{\tau}$. For fixed endpoints $a_0$ and $a_{\tau}$,
the probability that the path visits the given sequence of nodes reads
\begin{equation}
P(\gamma_{a_\tau a_0}) = P_{a_{\tau}a_{\tau-1}} \cdots P_{a_2a_1} P_{a_1a_0}.
\label{weight}
\end{equation}
The paths generated by the Markov chain representing the random walk
may thus have in general
different statistical weights, in contrast to the combinatorial problem where
each path is counted with the same weight, independently of the intermediate
nodes.

One case of much interest is generic (ordinary) random walk (GRW),
generalizing the Polya walk on a lattice,
which will be investigated in Section~\ref{grw}.
Neighbors are selected uniformly at each time step, \ie, $P_{ba} = A_{ba}/k_a$,
and hence
\begin{equation}
P(\gamma_{a_\tau a_0}) = \prod_{i=0}^{\tau-1} \frac{1}{k_{a_i}},
\label{weight2}
\end{equation}
which is the same weight as in (\ref{GFT}). Therefore, the trajectories
entering the field-theoretical derivation of path integrals are the same as
those generated by GRW.
For GRW on $k$-regular graphs,
there is a simple correspondence between the combinatorial result (\ref{Nba})
and the
probabilistic one~(\ref{pba}). One can interpret $N_{ba}(n)/k^n$
as the probability $p_{ba}(\tau)$ of reaching node $b$ after $\tau=n$ steps,
starting from $a$, since the numerator $N_{ba}(n)$ is the number
of paths of length $n$ between $a$ and $b$, while the denominator $k^n$ is the
number of paths of length $n$ starting from $a$ and ending anywhere.

A natural question arises:
{\it Can one find a stochastic matrix
which generates trajectories between given endpoints with uniform weights,
irrespective of intermediate nodes, for an arbitrary (non-regular) graph?}
In other words:
{\it Is there a random walk such that
all trajectories between two given endpoints are equally probable?}
A positive answer to the above question is provided by maximal entropy random
walk (MERW), to be investigated in Section~\ref{merw}.

Before this, let us recall some basic properties of the Markov chain
defined by the transition matrix $P$.
As already mentioned, we will restrict ourselves to the situation where
the transition matrix is primitive.
In this case, the probability distribution $p_{ba}(\tau)$
tends to a limiting distribution $\lim_{\tau\rightarrow \infty} p_{ba}(\tau) =
\pi_b$, independently of the initial point $a$. This unique distribution is
given by the normalized left eigenvector of the stochastic matrix corresponding
to unit eigenvalue:
\begin{equation}
\pi_b = \sum_a P_{ba} \pi_a,\qquad \sum_a \pi_a=1.
\label{stationary}
\end{equation}

\subsection{The entropy of a random walk}

Let us denote by $P(a_\tau,\ldots,a_1,a_0)$
the probability that a random walker follows a path
$a_0 \rightarrow a_1 \rightarrow \ldots \rightarrow a_\tau$.
The Markov property expressed by the master equation (\ref{master})
implies that this probability reads
\begin{equation}
P(a_\tau,\ldots,a_1,a_0) = P_{a_\tau a_{\tau-1}} \ldots P_{a_2a_1}P_{a_1a_0}
\pi_{a_0}(0),
\label{ppath}
\end{equation}
where $\pi(0)$ is the probability distribution of the initial point.

We define the entropy $H_\tau$ of the ensemble of paths of length $\tau$ as:
\begin{equation}
H_\tau = - \sum_{a_0,a_1,\ldots,a_\tau} P(a_\tau,\ldots,a_1,a_0)
\ln P(a_\tau,\ldots,a_1,a_0),
\label{Shannon}
\end{equation}
where the sum effectively runs over all the allowed paths of length $\tau$
generated by the master equation (\ref{master}).
Inserting Eq.~(\ref{ppath}) into (\ref{Shannon}), we find that
the entropy is asymptotically produced at a constant rate,
$h_\RW=\lim_{\tau\to\infty}(H_\tau/\tau)$, which reads
\begin{equation}
h_\RW= -\sum_{a} \pi_a \sum_b P_{ba} \ln P_{ba},
\label{h}
\end{equation}
independently of the initial distribution $\pi(0)$.
This means that after a long time, when the process
has reached its stationary state,
the entropy production rate is equal to a local production rate,
$h_a = - \sum_b P_{ba} \ln P_{ba}$, averaged over $\pi_a$.
The stationary distribution $\pi_a$ is
itself entirely determined by the stochastic matrix $P$ defining the random
walk
(see Eq.~(\ref{stationary})), and so is the entropy production rate (\ref{h}).

In the following we shall compare the entropies of two different types of
random walk:
generic random walk (GRW), which locally maximizes entropy production,
as already mentioned earlier,
and maximal entropy random walk (MERW), which maximizes entropy globally.

\subsection{Generic random walk (GRW)}
\label{grw}

\begin{figure}[!ht]
\begin{center}
\includegraphics[width=6.5cm]{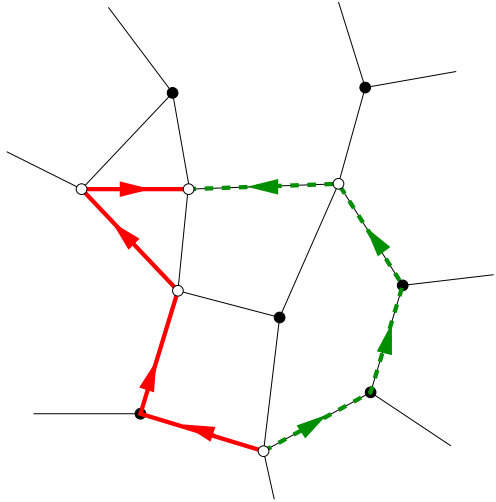}
\end{center}
\caption{\label{fig2}
(Color online).
Two paths (solid and dashed line) on a graph with
the same endpoints and length, but different statistical weights
in GRW, because they pass through different nodes.
Skipping probabilities attached to the endpoints,
the solid path has probability $1/4\cdot 1/3\cdot 1/4 \cdot 1/4=1/192$,
whereas the dashed one has probability $1/4\cdot 1/3\cdot 1/3 \cdot
1/4=1/144$.}
\end{figure}

Generic random walk (GRW),
generalizing the Polya walk on a lattice,
has already been defined above Eq.~(\ref{weight2}).
The particle sitting at node $a$ with degree $k_a$
hops to one of the neighboring nodes $b$
without giving a preference to any of them,
\ie, with probability $1/k_a$, so that
\begin{equation}
P_{ba} = \frac{A_{ba}}{k_a}.
\label{Pg}
\end{equation}
This maximally random choice at each time step corresponds
to a local maximization of the entropy production.
The local entropy $h_a = - \sum_b P_{ba} \ln P_{ba}$ is indeed maximized for
the uniform selection of neighbors (\ref{Pg}).
One may ask whether this choice also maximizes
the entropy production rate $h_\RW$~(\ref{h}) of entire paths.
We expect that it does not, because the statistical weights of paths
for GRW are given by Eq.~(\ref{weight2}) and hence paths are not equally probable in general,
even if they have identical length and endpoints (see Figure~\ref{fig2}).

The stationary distribution $\pi_a$ obtained from (\ref{stationary}) for GRW
reads
\begin{equation}
\pi_a = \frac{k_a}{2L}.
\label{ssg}
\end{equation}
The stationary probabilities of GRW are {\it directly} proportional to the node
degrees.
This is to be contrasted with Eq.~(\ref{weight2}),
which states that the contribution of intermediate nodes to the weight
of a given path is {\it inversely} proportional to their degrees.

Inserting Eq.~(\ref{ssg}) into (\ref{h}), we find the entropy production rate
$h_\GRW$
of generic random walk in the form:
\begin{equation}
h_\GRW = \frac{1}{2L}\sum_a k_a \ln k_a.
\label{hgrw}
\end{equation}

\subsection{Maximal entropy random walk (MERW)}
\label{merw}

We now turn to the question raised in Section~3.1.
If there is a random walk which maximizes entropy,
we expect that all paths of given length between two given endpoints
will be equally probable.
The weights of paths may however still depend on the endpoints,
as the contribution of endpoints will eventually disappear
in the limit $\tau \rightarrow \infty$, as it did in Eq.~(\ref{slambda}).
We are therefore looking for a stochastic matrix $P$
such that the weight (\ref{weight}) is a function of length $\tau$ and
endpoints $a,b$ only.
Moreover, as we expect that the matrix $P$ should have entropy
$h=\ln\lambda_1$,
the construction must somehow be related to the largest eigenvalue of the
adjacency matrix.

It can be checked that maximal entropy random walk (MERW)~\cite{ref:us},
defined by the following transition probabilities:
\begin{equation}
P_{ba} = \frac{A_{ba}}{\lambda_1} \frac{\psi_{1b}}{\psi_{1a}},
\label{Pmax}
\end{equation}
fulfills all the requirements.
First, $P$ is a stochastic matrix.
The components $\psi_{1a}$
of the eigenvector corresponding to the largest eigenvalue $\lambda_1$
are all positive, by virtue of the Perron-Frobenius theorem.
One can check that the row sums are equal to unity: $\sum_{b} P_{ba}=1$.
Finally, MERW conforms with the graph structure.

The stationary distribution (\ref{stationary}) for MERW
can be checked to be given by the squared components
of the normalized eigenvector $\psi_1$:
\begin{equation}
\pi_a = \psi_{1a}^2,
\label{psi2}
\end{equation}
so that the L$^1$ normalization (\ref{stationary}) of the $\pi_a$
nicely matches the L$^2$ normalization (\ref{Apsi}) of the $\psi_{ia}$.
Inserting the above result into (\ref{h}),
we obtain an expression for the entropy production rate,
\beq
h_\MERW = \ln \lambda_1=h, \label{hmerw}
\eeq
which indeed coincides with the combinatorial entropy of paths (\ref{slambda}).
It can also be checked that the statistical weight (\ref{weight}) for a path
$\gamma_{a_\tau a_0} = (a_\tau,\ldots,a_1,a_0)$ is independent
of intermediate nodes. Indeed, inserting~(\ref{Pmax}) into (\ref{weight}), we
obtain
\begin{equation}
P(\gamma_{a_\tau a_0}) = \frac{1}{\lambda_1^\tau}
\frac{\psi_{1a_\tau}}{\psi_{1a_0}}.
\label{ww}
\end{equation}
This expression only depends (exponentially) on the path length $\tau$,
and on the endpoints $a_0$ and $a_\tau$, through the components of $\psi_1$.
This means that all paths of length $\tau$ from $a_0$ to $a_\tau$ are indeed
equally probable.
In other words, the probability measure on this ensemble of paths is uniform,
and the corresponding entropy is maximal.

To the best of our knowledge,
MERW has been introduced and studied for the first time in our recent
work~\cite{ref:us}
in the context of random walk and path integrals.
However, the construction of stochastic processes with maximal entropy
in the framework of information theory is much older,
as it dates back to Shannon~\cite{ref:shannon}.
The concept has been formalized by Parry~\cite{ref:parry} as intrinsic
Markov chains.
In the context of ergodic theory,
the stationary distribution (\ref{psi2}) is referred to as
a Shannon-Parry measure~\cite{ref:brin},
whereas yet other related matters are discussed in Refs.~\cite{ref:hetherington}
and~\cite{ref:neil}.

The Perron-Frobenius eigenvector $\psi_1$ can be interpreted as the ground state of the
Hamiltonian
$H=(H_{ab})$, with $H_{ab} = k_\max \delta_{ab} -
A_{ab}=-\Delta_{ab}+V_a\delta_{ab}$,
where $V_a=k_\max-k_a$, and $k_\max$ is the maximal node degree in the graph.
In other words, $\psi_1$ is the ground-state wavefunction of the tight-binding
equation
\begin{equation}
(H \psi_i)_a = (-\Delta \psi_i)_a + V_a\psi_{ia} = E_i \psi_{ia},
\label{Hpsi}
\end{equation}
with
\beq
\quad V_a=k_\max-k_a,\quad E_i=k_\max-\lambda_i,
\eeq
and the stationary distribution (\ref{psi2})
is the square of this ground-state wavefunction.
The potential $V_a$ is non-negative.
It is positive (\ie, repulsive) for nodes whose degree is smaller than $k_\max$
--- the smaller the degree, the larger the repulsion.
All the eigenvalues $E_i$ of $H$ are clearly non-negative,
the ground-state eigenvalue $E_1 = k_\max - \lambda_1$ being the smallest one.
For a $k$-regular graph, the potential vanishes identically.
The Hamiltonian~$H$ thus describes the propagation of a free particle.
The ground-state energy is $E_1=0$ and we have $\psi_{1a} = 1/\sqrt{N}$,
so that $\pi_a=1/N$, where $N$ is the number of nodes of the graph.
Hence the stationary measure $\pi_a$ is uniform over the $k$-regular graph.

\subsection{Effective degrees}

It is interesting to illustrate the above considerations
by associating effective degrees to the entropies of GRW and of MERW on a
graph.
Consider an arbitrary finite graph whose $N$ nodes ($a=1,\dots,N$) have degrees
$k_a$.
We introduce the following three definitions of its effective degree:

\begin{itemize}

\item
The first effective degree of a graph is simply its mean degree,
\beq
K_1=\frac{1}{N}\sum_ak_a=\frac{2L}{N}.
\label{k1}
\eeq
The node degrees $k_a$ indeed sum up to twice the number of links $L$
(see (\ref{traces})).

\item
The second definition is the GRW-based degree $K_2=\exp(h_\GRW)$,
where the entropy $h_\GRW$ of GRW is given by (\ref{hgrw}), hence
\beq
K_2=\exp\left(\frac{1}{2L}\sum_ak_a\ln k_a\right)
=\left(\prod_ak_a^{k_a}\right)^{1/(2L)}.
\label{k2}
\eeq

\item
The third one is the MERW-based degree $K_3=\exp(h_\MERW)$,
where the entropy $h_\MERW$ of MERW is given by (\ref{hmerw}), hence
\beq
K_3=\lambda_1.
\label{k3}
\eeq

\end{itemize}

The intuition behind the entropic definitions of the effective
node degrees $K_2$ and $K_3$ is the following.
On a $k$-regular graph, the number of paths of length $n$ starting from
a given node grows as $N(n)=k^n$, so that the degree $k$ is related to the
entropy production rate as $k=\exp(h)$.
A natural extension of this relation to irregular graphs leads to (\ref{k2}) or
(\ref{k3}) if one uses the local or global rule for maximal entropy production,
respectively.

The three definitions of the effective degree yield in general different values.
Their number-theoretical natures are very different:
$K_1$ is a rational number,
whereas $K_2$ is a fractional power of an integer,
and $K_3$ is a solution of an algebraic equation of degree at most $N$.
The effective degrees obey the inequalities
\beq
k_\min\le K_1\le K_2\le K_3\le k_\max,
\label{kineqs}
\eeq
where $k_\min$ and $k_\max$ are the minimal and maximal values of the node
degrees.
The first and fourth of these inequalities are obvious,
whereas the third one just expresses that MERW indeed has maximal entropy,
and the second one originates in the convexity of the free energy
\beq
F(\beta)=\ln\sum_ak_a^\beta.
\eeq
We have indeed $\ln K_2=F'(1)\ge\ln K_1=F(1)-F(0)$.

\section{Stochastic quantization}
\label{sec4}

As already mentioned, the trajectories generated by a random
walk can be used to define quantum amplitudes. In this section we
make this statement more precise within the framework of stochastic
quantization
(see, \eg,~\cite{ref:zinn}).
As one can anticipate,
MERW will reproduce the path-integral propagator~(\ref{GFPI}),
and GRW the field-theoretical propagator (\ref{GFT}).

Let us start with MERW.
We consider paths generated by the stochastic matrix (\ref{Pmax}),
and restrict ourselves to equilibrium paths, initiated from the stationary
state (\ref{psi2}).
The probability of such a path is
\begin{equation}
P_\eq(\gamma_{a_\tau a_0}) = P(a_\tau,\ldots,a_1,a_0) \pi_{a_0}
= \e^{- \tau h} \psi_{1a_\tau} \psi_{1a_0},
\end{equation}
with $h=\ln\lambda_1$, as one can see by multiplying (\ref{ww}) by $\psi_{1a_0}^2$.
Keeping endpoints $a_0=a$ and $a_\tau=b$ fixed and summing
over intermediate states $a_1,\ldots, a_{\tau-1}$, we obtain
a sum over all paths $\gamma_{ba}(\tau)$ from $a$ to $b$ of length $\tau$:
\begin{equation}
\sum_{\{\gamma_{ba}\}} P_\eq(\gamma_{ba}) =
\sum_{a_{\tau-1}\ldots a_1} P_\eq(b,a_{\tau-1},\ldots,a_1,a) =
N_{ba}(\tau) \e^{-\tau h} \psi_{1b} \psi_{1a}.
\end{equation}
If we now multiply both sides by an additional exponential weight
$\e^{-\hat{\mu} \tau}$
with a positive parameter $\hat{\mu} > 0$,
and sum over all integer values of $\tau$ from zero to infinity, we eventually
obtain
\begin{equation}
\sum_\tau\e^{-\hat{\mu} \tau} \sum_{\{\gamma_{ba}(\tau)\}}
P_\eq(\gamma_{ba}(\tau)) = \psi_{1b} G_{ba} \psi_{1a},
\label{ewf}
\end{equation}
where $G_{ba}$ is the propagator derived in the path-integral formalism
(\ref{laplace}) with $\mu = \hat{\mu} + h$.
The above expression becomes singular as the sum diverges
for $\hat{\mu} \rightarrow 0$, \ie, $\mu\to h=\ln\lambda_1$.
This situation describes the limit of a massless particle,
in which
all paths of any length are equally probable. The right-hand side
of Eq.~(\ref{ewf}) has a characteristic sandwich form where $G_{ab}$
stands between wave functions representing external states $\psi_{1a} =
\sqrt{\pi_a}$ (see (\ref{psi2})).

We can now repeat the same construction for GRW with the stochastic matrix
(\ref{Pg}).
As in the previous case, it is convenient to define a state function $\hat\psi$
as the square root of the stationary probability (\ref{ssg}):
$\hat{\psi}_a = \sqrt{\pi_a}=\sqrt{k_a/(2L)}$.
The probability of an equilibrium path reads
\begin{equation}
P_\eq(\gamma_{a_\tau a_0}) = P(a_\tau,\ldots,a_1,a_0) \pi_{a_0} =
\hat{\psi}_{a_\tau} \hat{\psi}_{a_0}
\sqrt{k_{a_\tau} k_{a_0}} \prod_{i=0}^{\tau}\frac{1}{k_{a_i}},
\end{equation}
as one can see by multiplying (\ref{weight2}) by $(\ref{ssg})$.
Applying the same procedure as for MERW, we obtain
\begin{equation}
\sum_{\tau=0}^\infty \e^{-\hat{\mu} \tau} \sum_{\{\gamma_{ba}(\tau)\}}
P_\eq(\gamma_{ba}(\tau)) = \hat{\psi}_{b} \widehat{G}_{ba} \hat{\psi}_{a},
\label{newf}
\end{equation}
where $\widehat{G}_{ba}$ is equal to the field-theoretical propagator
(\ref{GFT}). As already mentioned, the two propagators are identical only
for $k$-regular graphs. In this case, the largest eigenvalue
of the adjacency matrix is $\lambda_1=k$, the entropy rate is $h=\ln k$,
$\mu = \hat{\mu} + \ln k$, $\hat{\psi}_a = \psi_{1a} = \sqrt{k/(2L)}$,
and the stochastic matrix of MERW (\ref{Pmax}) is identical to that of GRW
(\ref{Pg}).

\begin{figure}[!ht]
\psfrag{t}{$t$}\psfrag{a}{a}\psfrag{b}{b}\psfrag{c}{c}
\begin{center}
\includegraphics[width=7cm]{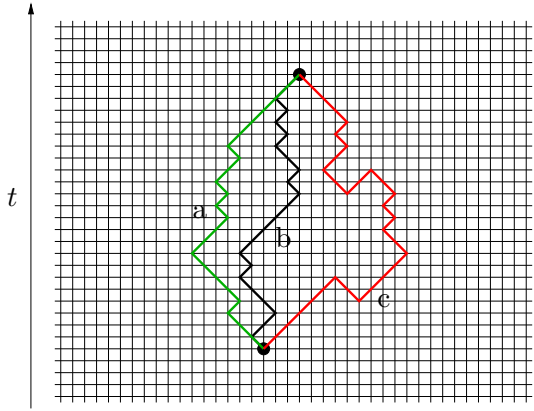}
\end{center}
\caption{\label{fig4}
(Color online).
Trajectories of a non-relativistic particle on a (1+1)-dimensional square
lattice
form directed polymers (a,b), whereas trajectories which turn back in time are
excluded (c). Both (a) and (b) have the same length and hence they are equally probable within the MERW (path-integral) formalism.}
\end{figure}

Let us conclude this section with a remark on the relationship with
non-relativistic Quantum Mechanics.
In the relativistic case discussed so far, the parameter
$\tau$ was treated as {\it proper} time. The Euclidean propagators, either in the
path-integral approach (\ref{ewf}) or in the field-theoretical formalism (\ref{newf}),
were independent of $\tau$, since the latter variable was summed over.
One can, however, treat $\tau$ as a universal time in a non-relativistic spacetime,
and view the graph (or lattice) as a discretization of space only, hence
skipping the summation over $\tau$.
This leads to a non-relativistic propagator for $(d+1)$-dimensional QM:
\begin{equation}
\sum_{\{\gamma_{ba}(\tau)\}} P_\eq(\gamma_{ba}(\tau)) =
\psi_{1b} K_{ba}(\tau) \psi_{1a},
\end{equation}
where $K_{ba}(\tau)$ (or alternatively $\widehat{K}_{ba}(\tau)$)
is defined as a sum over trajectories of fixed length $\tau$.
The non-relativistic propagator is thus
related to the Euclidean relativistic propagator by a Laplace transform:
\begin{equation}
G_{ba}(\hat{\mu})=\sum_\tau K_{ba}(\tau)\e^{-\hat{\mu}\tau}.
\end{equation}
One can also explicitly add a time direction to the discretized theory by
stacking $d$-dimensional lattices on top of each other. Doing so,
one obtains a $(d+1)$-dimensional foliated lattice whose time slices
are identical clone copies of $d$-dimensional space.
In this context, the trajectories of a particle form directed polymers,
which never go backward in time (see Figure~\ref{fig4}).
Thus MERW generates maximally entropic directed polymers
which are equally probable in an ensemble of polymers having fixed endpoints.
This is not the case for directed polymers generated by GRW
in a curved background, discretized as a non-regular lattice.

\section{Examples of finite graphs}
\label{sec5}

In what follows we discuss properties of the stationary state for GRW and MERW on various finite graphs.

\subsection{$k$-regular graphs}

As already mentioned, the simplest situation is that of a $k$-regular graph,
where all the nodes have the same degree $k$.
Both GRW and MERW are defined by the uniform transition probabilities
$P_{ba}=1/k$ if nodes $a$ and $b$ are neighbors.
The two processes therefore coincide, no matter how complicated the topology of the graph is.
The corresponding stationary distribution is uniform: $\pi_a=1/N$ for all
nodes.
We have consistently
\beq
K_1=K_2=K_3=k.
\eeq

\subsection{Bipartite graphs}
\label{sec:bip}

The next case, in order of increasing complexity, is that of bipartite graphs.
We will consider the class of finite bipartite graphs
such that all nodes in one subset of the graph
have identical degree $k_1$, while in the other subset all nodes have degree $k_2$.
The numbers of nodes of each type are then $N_1=L/k_1$ and $N_2=L/k_2$,
where $L$ denotes the number of links.
For both GRW and MERW, the transition probabilities take two values:
$P_{21}=1/k_1$ (if node $a$ has degree~$k_1$ and node $b$ has degree $k_2$)
and $P_{12}=1/k_2$ (if node $a$ has degree $k_2$ and node $b$ has degree
$k_1$).
The two processes therefore again coincide.
The largest eigenvalue of the adjacency matrix is $\lambda=\sqrt{k_1k_2}$.
The corresponding eigenvector obeys $\psi_1:\psi_2=\sqrt{k_1}:\sqrt{k_2}$.
The stationary distribution takes two values:
$\pi_1=k_1/(2L)$ and $\pi_2=k_2/(2L)$.
The effective degrees read
\beq
K_1=\frac{2k_1k_2}{k_1+k_2},\qquad K_2=K_3=\sqrt{k_1k_2}.
\eeq
In other words, $K_1$ is the harmonic mean of both degrees,
whereas $K_2$ and $K_3$ coincide with their geometric mean.
These results hold irrespective of the size and topology of the graph.

The first non-trivial example of a bipartite but non-regular graph
corresponds to $k_1=2$ and $k_2=3$.
We have $K_1=12/5=2.4$, whereas $K_2=K_3=\sqrt{6}=2.449490$.
These two values are different, albeit very close to each other.

\subsection{The barred-square graph}

\begin{figure}[!ht]
\begin{center}
\includegraphics[angle=90,width=.2\linewidth]{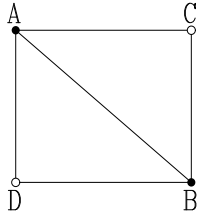}
\caption{\label{sq}
The barred-square graph.}
\end{center}
\end{figure}

The barred-square graph shown in Figure~\ref{sq} is the simplest example
of interest of a non-bipartite graph.

For GRW on this graph, the stationary distribution reads
\beq
\pi_\A=\pi_\B=\frac{3}{10},\qquad \pi_\C=\pi_\D=\frac{1}{5},
\eeq
so that we have $h_\GRW=(\ln 108)/5$.
The largest eigenvalue of the adjacency matrix is $\lambda=(\sqrt{17}+1)/2$.
The corresponding eigenvector obeys $\psi_A:\psi_C=(\sqrt{17}+1):4$.
Therefore, the stationary distribution for MERW reads
\beq
\pi_\A=\pi_\B=\frac{\sqrt{17}+1}{4\sqrt{17}}=0.310634,\quad
\pi_\C=\pi_\D=\frac{\sqrt{17}-1}{4\sqrt{17}}=0.189366.
\eeq
Finally, the effective degrees are
\beq
K_1=\frac{5}{2}=2.5,\quad K_2=108^{1/5}=2.550849,\quad
K_3=\frac{\sqrt{17}+1}{2}=2.561553.
\eeq
These three values are again very close to each other.

\subsection{Linear graphs}
\label{sec:lin}

\begin{figure}[!ht]
\begin{center}
\includegraphics[angle=90,width=.4\linewidth]{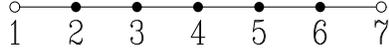}
\caption{\label{lin}
The linear graph with $N=7$ nodes.}
\end{center}
\end{figure}

Consider now the family of linear graphs made of $N\ge3$ nodes,
as shown in Figure~\ref{lin} for $N=7$.

The endpoints ($n=1$ and $n=N$) have degree 1,
whereas the $N-2$ inner points ($n=2,\dots,N-1$) have degree 2.
The mean degree thus reads
\beq
K_1=\frac{2(N-1)}{N}.
\eeq
For GRW on the linear graph, the stationary distribution is
\beq
\pi_1=\pi_N=\frac{1}{2(N-1)};\qquad \pi_n=\frac{1}{N-1}\qquad(n=2,\dots,N-1),
\label{dorw}
\eeq
and we have
\beq
K_2=2^{(N-2)/(N-1)}.
\eeq
For MERW the stationary distribution is
\beq
\pi_n=\frac{2}{N}\,\sin^2\frac{n\pi}{N+1}
\label{dmerw}
\eeq
and we have
\beq
K_3=\lambda=2\cos\frac{\pi}{N+1}.
\eeq

As the linear graph gets larger ($N\to\infty$),
the three effective degrees converge to the limiting value 2,
characteristic of the infinite chain, albeit at different rates:
\beq
K_1=2-\frac{2}{N},\quad K_2=2-\frac{2\ln 2}{N}+\cdots,\quad
K_3=2-\frac{\pi^2}{N^2}+\cdots
\eeq
The positive differences $K_2-K_1$, $K_3-K_2$, and $K_3-K_1$ are respectively
maximal
for $N=4$, 9, and 6.

\begin{figure}[!ht]
\begin{center}
\includegraphics[angle=-90,width=.5\linewidth]{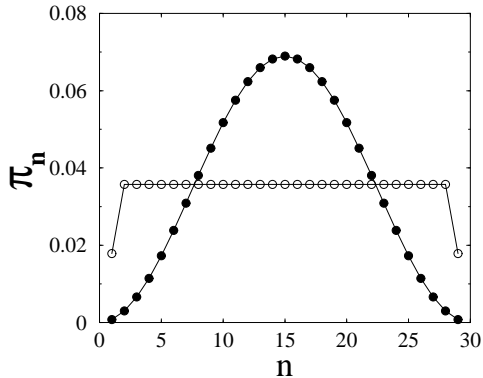}
\caption{\label{pi}
The stationary distributions (\ref{dorw}) of GRW (empty symbols)
and (\ref{dmerw}) of MERW (filled symbols)
on the linear graph with $N=29$ nodes.}
\end{center}
\end{figure}

Figure~\ref{pi} shows a comparison of the stationary distributions of GRW~(\ref{dorw})
and MERW (\ref{dmerw}) on the linear graph with $N=29$ nodes.
The effect of the endpoint impurities is strictly local in the case of GRW,
as it only affects the distribution at the endpoints themselves.
On the contrary, in the case of MERW we observe a non-local effect of
the endpoints
on the stationary distribution, which varies smoothly at the scale of the whole
graph.
Rather paradoxically, the MERW-based effective degree $K_3$ converges as
$1/N^2$,
whereas the other two have a slower linear convergence in $1/N$.

The above picture is generic.
Statistical properties of MERW may differ drastically from those of GRW.
We recall that the stationary distribution for MERW is
the square of the ground-state wavefunction of the Hamiltonian
involved in the tight-binding equation (\ref{Hpsi}).
The latter describes the motion of a QM particle in the presence of a repulsive potential
$V_a=k_\max-k_a$, supported by the impurity nodes whose degree is smaller than
$k_\max$. These impurities can be expected to generate strong, non-local
effects.

\section{One-dimensional examples: ladders}
\label{sec6}

\begin{figure}[!ht]
\begin{center}
\includegraphics[width=8cm]{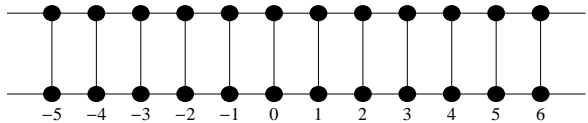}
\end{center}
\caption{\label{ladder}
The full ladder. Periodic boundary conditions along the chains are assumed.}
\end{figure}

We now turn to the investigation of MERW on extended structures,
starting from quasi one-dimensional systems --- ladder graphs.
The full graph consists of two symmetric closed chains of nodes connected by
rungs (see Figure~\ref{ladder}). It is a $3$-regular graph on which, as we have learned, GRW and MERW coincide.

In order to meet cases where MERW behaves in a non-trivial manner,
we remove some of the rungs~\cite{ref:us}.
Since the ground state is not degenerate and its wavefunction is expected to be symmetric
w.r.t.~the exchange of both lines of the ladder, it is sufficient
to consider the tight-binding equation~(\ref{Hpsi}) in the symmetric sector.
This equation takes the form
\beq
2\psi_{1a}-\psi_{1a-1}-\psi_{1a+1}+V_a\psi_{1a}=E_1\psi_{1a}, \label{tb1}
\eeq
where $E_1=3-\lambda_1$ and
\beq
V_a=\left\{\begin{array}{ll}
0 & \hbox{if the rung at position $a$ is present},\\
1 & \hbox{if the rung at position $a$ is absent}.
\end{array} \right.
\eeq
For definiteness we consider ladders whose length (circumference)
is an even number $2N$. We will index nodes in each line by
$a=-N,\ldots,N$, and impose periodic boundary conditions by identifying nodes
$N$ and $-N$.

\subsection{A single impurity}

\begin{figure}[!ht]
\begin{center}
{\hskip 28pt}\hspace{7mm}\includegraphics[width=8cm]{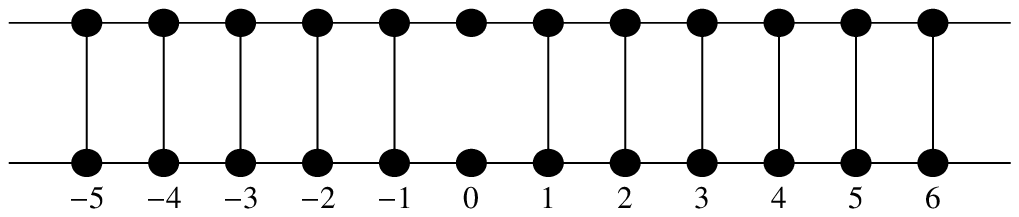}
\includegraphics[width=9cm]{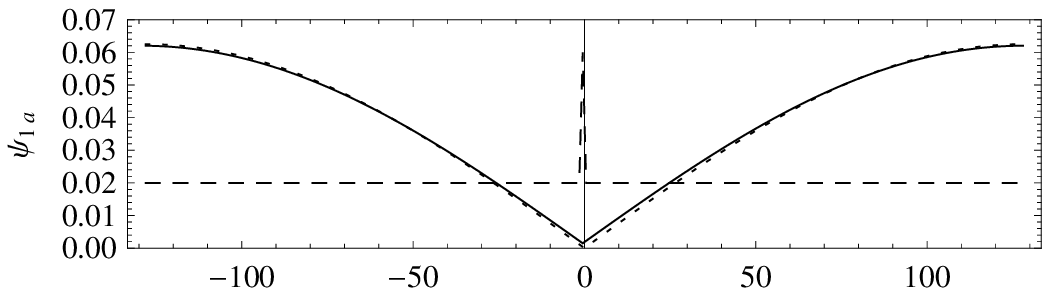}
\end{center}
\caption{\label{ladder2}
Top: a ladder with one rung at $a=0$ removed.
Bottom: Plot of the numerically obtained ground-state wavefunction $\psi_{1a}$
for $N=128$ (solid line)
compared with (\ref{sine1}) (dotted line).
The impurity potential $V_a$ is also shown (dashed line, rescaled).}
\end{figure}

Let us first create a single impurity at the origin
by removing the rung at node $a=0$ (see Figure~\ref{ladder2}).

The potential $V_a=\delta_{a0}$ is concentrated on the impurity.
Setting $E_1=2(1-\cos q)$, \ie, $\lambda_1=1+2\cos q$, with some unknown wavevector $q$,
the (unnormalized) wavefunction reads
\beq
\psi_{1a}=\cos((N-\abs{a})q).
\eeq
The matching condition on the impurity yields the quantization condition
\beq
2\sin q\,\tan(Nq)=1.
\label{cond1}
\eeq
The ground state corresponds to the smallest positive solution $q$ to the latter
equation. In the most interesting situation, namely for large $N$,
we have $q\approx\pi/(2(N+2))$, and hence
\begin{equation}
\psi_{1a} \approx \frac{1}{\sqrt{2N}}\ \sin\frac{(\abs{a}+2)\pi}{2N}.
\label{sine1}
\end{equation}
The presence of a single defect has a non-local effect on the stationary probability
distribution. The latter distribution is maximal at $a=N$, \ie,
at the node farthest from the removed rung, whereas it is minimal on the impurity.
Although the potential is concentrated at a single node,
it exerts a strong repulsion on a particle performing MERW.
The situation is somewhat similar to that of long linear graphs,
described in Section~\ref{sec:lin},
where the endpoints of the finite linear chain acted as impurities
whose effect was already non-local.

\subsection{Two impurities}

\begin{figure}[!ht]
\begin{center}
{\hskip 28pt}\hspace{7mm}\includegraphics[width=8cm]{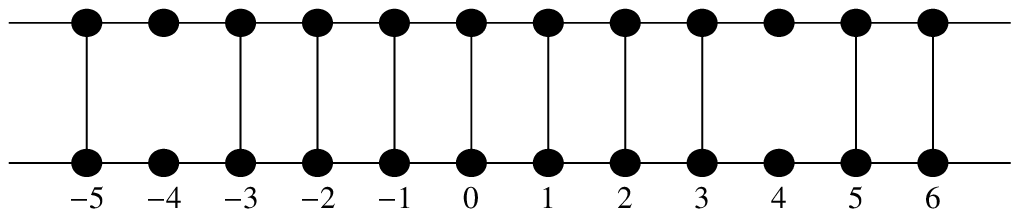}
\includegraphics[width=9cm]{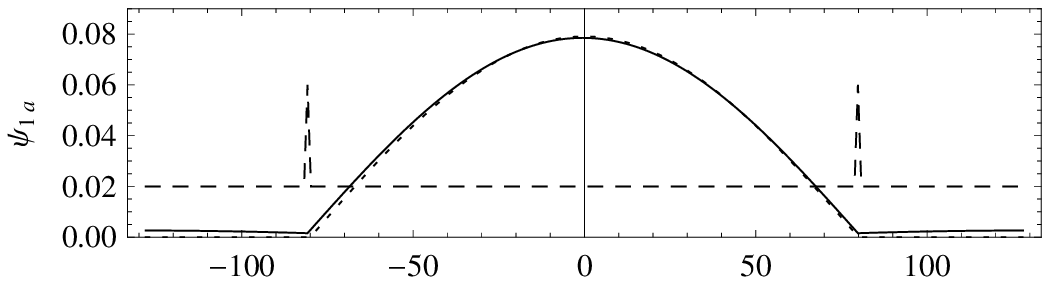}
\end{center}
\caption{\label{ladder3}
Top: a ladder with two rungs at $a=-n$ and $a=n$
removed, for $n=4$. Bottom: Plot of the numerically obtained ground-state
wavefunction
$\psi_{1a}$ for $N=128$ and $n=80$ (solid line) compared with (\ref{sine2})
(dotted line).
The impurity potential $V_a$ is also shown (dashed line, rescaled).}
\end{figure}

We now consider what happens when two rungs are removed
at positions $a=n$ and $a=-n$ (see Figure~\ref{ladder3}).

The impurity potential $V_a=\delta_{a,n}+\delta_{a,-n}$
and the ground-state wavefunction are again symmetric in $a$.
The (unnormalized) wavefunction reads
\beq
\psi_{1a}=\left\{\matrix{
\cos((N-n)q)\,\cos(aq)\quad\hfill&(-n\le a\le n),\hfill\cr
\cos((N-\abs{a})q)\,\cos(nq)\hfill&(n\le\abs{a}\le N).
}\right.
\eeq
The matching condition on the impurity yields the quantization condition
\beq
\sin q\left(\tan(nq)+\tan((N-n)q)\right)=1.
\label{cond2}
\eeq

For large $N$ and $n$, the smallest solution reads approximately
\beq
q_1\approx\min\left(\frac{\pi}{2n},\frac{\pi}{2(N-n)}\right).
\eeq
The ground-state wavefunction, and therefore the stationary distribution of
MERW,
essentially live on the larger part of the ladder which is free of defects.
For $n>N/2$, this larger region is the central one.
The ground-state wavefunction is well approximated by
\begin{equation}
\psi_{1a} \approx \left\{
\begin{array}{ll}
\frad{1}{\sqrt{2n}}\,\cos\frac{a\pi}{2n} & (-n\le a\le n), \label{sine2} \\
0 & \ {\rm otherwise.}
\end{array} \right.
\end{equation}

\subsection{A single attractive impurity}

\begin{figure}[!ht]
\begin{center}
{\hskip 22pt}\hspace{7mm}\includegraphics[width=8cm]{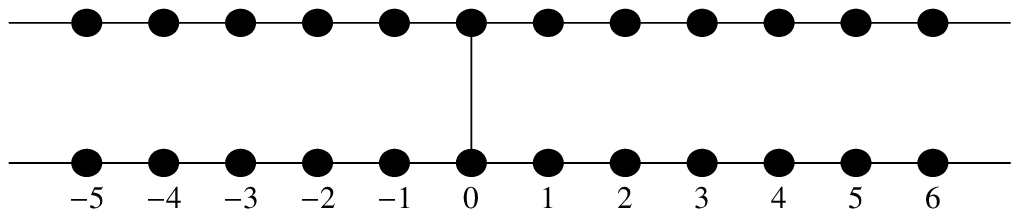}
\includegraphics[width=9cm]{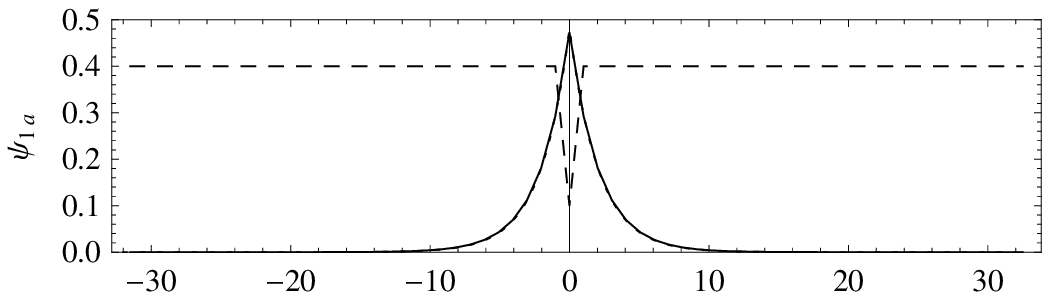}
\end{center}
\caption{\label{ladder4}
Top: A ladder with a single rung at $a=0$.
Bottom: Plot of the numerically obtained ground-state wavefunction $\psi_{1a}$
for $N=32$
(solid line) compared with (\ref{exp1}) (dotted line, hardly visible).
The impurity potential $V_a$ is also shown (dashed line, rescaled).}
\end{figure}

One can also consider the reverse situation of attractive defects,
\ie, nodes having a degree larger than average.
The simplest example is provided by a ``degenerate ladder'' with a single rung at $a=0$
(see Figure~\ref{ladder4}).
The nodes at the endpoints of the rung have degree $k=3$,
whereas all other nodes have $k=2$.
In this case the impurity potential reads $V_a=1-\delta_{a0}$.
It is everywhere repulsive, except at the origin.
The ground state can thus be expected to be localized around the origin.

This is indeed what happens.
Setting $E_1=3-2\cosh\theta$, \ie, $\lambda_1=2\cosh\theta$,
the (unnormalized) wavefunction reads
\beq
\psi_{1a}=\cosh((N-\abs{a})\theta).
\eeq
The matching condition on the impurity yields the quantization condition
\beq
2\sinh\theta\,\tanh(N\theta)=1.
\label{cond3}
\eeq
The ground state is the unique bound state,
corresponding to the real solution $\theta_1$ to the latter equation.
In the situation of most interest, namely~$N$ large,
we obtain a non-trivial limiting solution so that $\sinh\theta_1=1/2$,
\ie, $\theta_1=\ln((\sqrt{5}+1)/2)$, and hence $\lambda_1=\sqrt{5}$.
The limiting normalized ground-state wavefunction reads
\begin{equation}
\psi_{1a}=\frac{\e^{-|a|\theta_1}}{\sqrt{\tanh\theta_1}}.
\label{exp1}
\end{equation}
The particle performing MERW is localized in the vicinity of the attractive
defect (rung).
The stationary probability $\pi_a=\psi_{1a}^2$
of finding the particle at a distance $a$ from the defect
falls off exponentially as $\exp(-2|a|\theta_1)$.
The corresponding localization length is $\xi=1/\theta_1=2.078087$.

The existence of a localized ground state implies that
the effective degree $K_3=\lambda_1=\sqrt{5}=2.236068$
remains strictly larger than $K_1=K_2=2$, even in the limit of an
infinitely long system.

\subsection{Diluted ladders}
\label{sec:lif}

We now consider the disordered situation
where impurities are distributed at random all over the ladder
with some concentration $q$.
In other words, we consider a diluted ladder graph where rungs are removed,
independently of each other, with probability $q$.
The tight-binding equation for the ground state is given by (\ref{tb1}),
where the potential $V_a$ is a sequence of i.i.d.~variables with the binary
distribution
\beq
V_a=\left\{\begin{array}{ll}
0 & \hbox{with probability $p$},\\
1 & \hbox{with probability $q=1-p$.}
\end{array} \right.
\eeq

The problem therefore maps onto the one-dimensional tight-binding Anderson
model
with diagonal disorder.
Generic features of this model, such as the density of states or the
localization length,
have been investigated at length.
In the present situation,
we are however chiefly interested in the ground state of the model
on a finite system of length $L$.
This question is related to the behavior of the density of states $\rho(E)$
near the bottom of the spectrum ($E\to0$).
The latter is known to have the form
of an exponentially small Lifshitz tail~\cite{ref:lif}.
Lifshitz tails have been studied extensively,
both from a mathematically rigorous standpoint~\cite{ref:mplif}
and in various physical contexts, including in particular the diffusion
of particles in the presence of randomly placed absorbing traps~\cite{ref:trap}.

In the one-dimensional case, the Lifshitz argument goes as follows.
Low-lying states are expected to live on long wells,
\ie, ordered sequences of sites without impurities.
Assume there is a well between sites~$b$ and $c$, \ie, $V_b=V_c=1$, whereas $V_a=0$
for $b<a<c$.
There will be an eigenstate living essentially on that well,
and resembling the free mode on the well with Dirichlet boundary conditions
at $b$ and $c$, \ie,
$\psi_a\sim\sin((a-b)\pi/\ell)$, where $\ell=c-b$ is the length of the well.
The corresponding energy is $E\approx\pi^2/\ell^2$.
This line of thought can be used to estimate the fall-off of
the density of states near the bottom of the spectrum ($E\to0$).
Long wells ($\ell\gg1$) occur in the chain with an exponentially small
probability of order~$p^\ell$.
Eliminating~$\ell$ for the corresponding energy $E$, we thus obtain
\beq
\rho(E)\sim\exp\left(-\frac{\pi\abs{\ln p}}{\sqrt{E}}\right).
\label{lifrho}
\eeq
In the one-dimensional situation the above result is virtually exact,
up to absolute prefactors involving oscillating functions~\cite{ref:lifone}.
The same picture can be used to estimate the ground-state
energy~$E_1$ on a finite system.
According to a well-known argument of extreme-value statistics,
the typical length $\ell_1$ of the longest well on a finite system of length $L$
is such that~$p^{\ell_1}$ is of order $1/L$.
We therefore predict that the ground state is typically localized
in a Lifshitz well of length
\beq
\ell_1\approx\frac{\ln L}{\abs{\ln p}},
\label{lifl}
\eeq
and that the corresponding energy reads
\beq
E_1\approx\left(\frac{\pi\abs{\ln p}}{\ln L}\right)^2.
\label{life}
\eeq

\begin{figure}[!ht]
\begin{center}
\includegraphics[width=12cm]{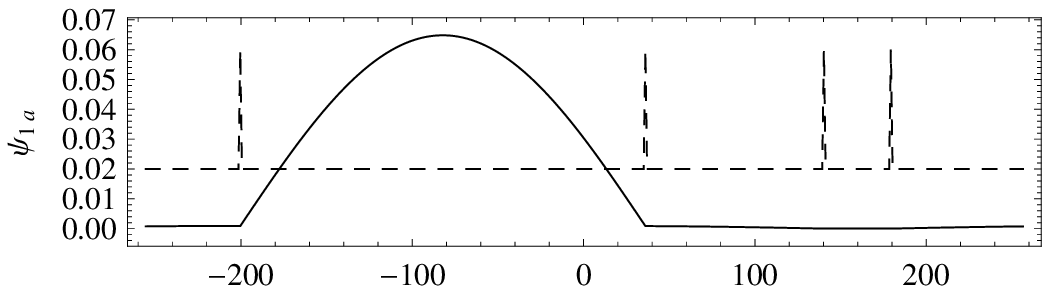}
\includegraphics[width=12cm]{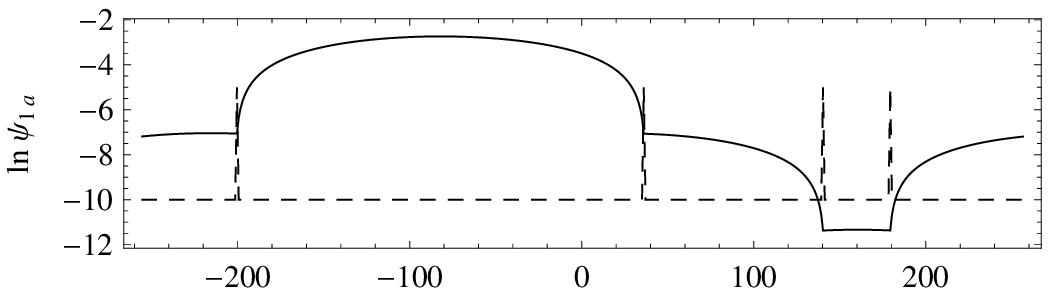}
\end{center}
\caption{\label{fig6a}
Top: stationary distribution $\pi_a$ of MERW
on one configuration of the disordered ladder for
a size $L=512$ and a density $q=0.01$ of defects.
Bottom: same data plotted on a logarithmic scale.
The impurity potential $V_a$ is also shown (dashed line, rescaled).}
\end{figure}

\begin{figure}[!ht]
\begin{center}
\includegraphics[width=12cm]{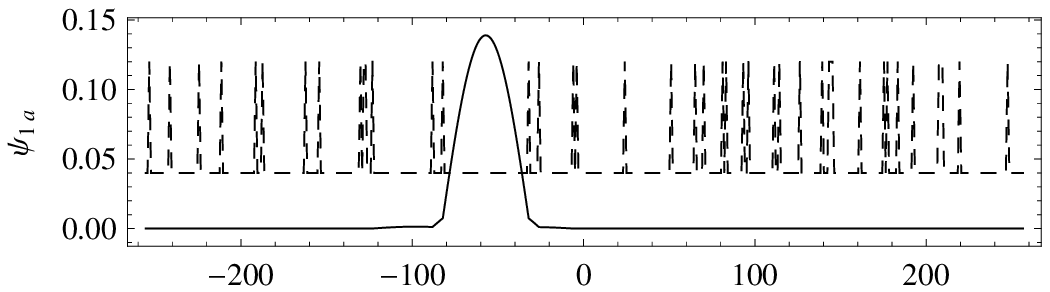}
\includegraphics[width=12cm]{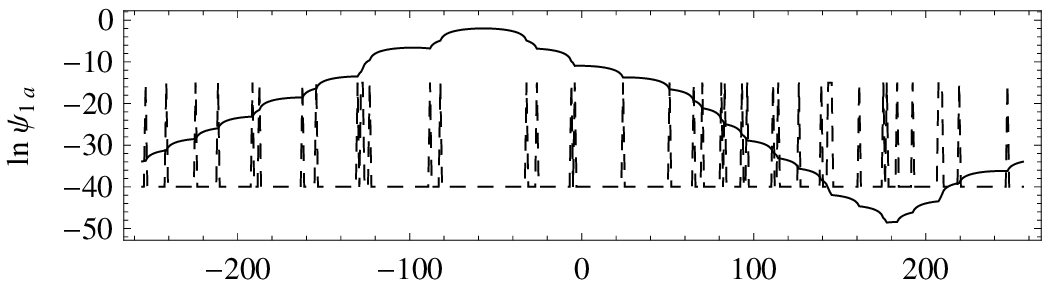}
\end{center}
\caption{\label{fig6b}
Same as Figure~\ref{fig6a}, for a density of defects $q=0.1$.}
\end{figure}

Figures~\ref{fig6a} and~\ref{fig6b} show linear and logarithmic plots of
the stationary distribution of MERW on disordered ladders,
obtained by a numerical diagonalization of the adjacency matrix for a system
size $L=512$.
For $q=0.01$ (see Figure~\ref{fig6a}), the estimate (\ref{lifl}) yields
$\ell\approx620$.
Both length scales $L$ and $\ell$ are comparable,
in agreement with the data showing a macroscopically large ``dome''.
For $q=0.1$ (see Figure~\ref{fig6b}), we obtain $\ell\approx59$.
Disorder is strong enough to observe the Lifshitz phenomenon,
\ie, the localization of the stationary distribution in the longest region
without defects,
which extends approximately between the positions $-80$ and $-30$,
so that its length is indeed of order 50.
The usual localization effect, \ie, the exponential fall-off of the
wavefunction over a characteristic length given by the localization length
$\xi$,
is also clearly visible on the lower panel of Figure~\ref{fig6b}.

\begin{figure}[!ht]
\begin{center}
\includegraphics[angle=-90,width=.5\linewidth]{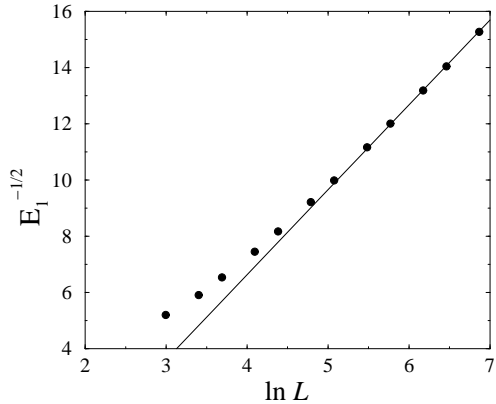}
\end{center}
\caption{\label{gseplot}
Plot of the inverse square root of the mean ground-state energy
on disordered ladders against $\ln L$, for sizes $L=20,\dots,960$ and $q=0.1$.
The solid line with slope $1/(\pi\abs{\ln 0.9})$ and a fitted intercept
corroborates the estimate (\ref{life}).}
\end{figure}

Figure~\ref{gseplot} shows a plot of $E_1^{-1/2}$ against $\ln L$,
for a concentration of defects $q=0.1$, and system sizes $L$ ranging from 20 to 960.
The ground-state energy has been obtained by diagonalizing numerically the adjacency
matrix, and averaging the outcome over many disorder configurations.
The data fully confirm the Lifshitz prediction (\ref{life}).
Finally, let us mention that the effective degrees
$K_1=3-q$ and $K_2\approx3-2q\ln(3/2)\approx3-0.810930\,q$
depart linearly from the value 3
for an infinite ladder with a weak concentration $q$ of defects,
whereas $K_3=3-E_1$ still goes asymptotically to~3,
albeit with a logarithmic finite-size correction (see~(\ref{life})).

\subsection{The Fibonacci ladder}
\label{sec:fib}

We close this section on ladders by considering a deterministic but non-periodic
configuration of the removed rungs.
We choose for definiteness the Fibonacci sequence 0100101001001010010100101001001\dots
This infinite sequence can be built in two alternative ways:
either recursively, or by an explicit formula~\cite{ref:concrete}.
In the first method, the sequence is built as the fixed point of the substitution:
\beq
\sigma:\left\{\matrix{0\to01\cr1\to0\hfill},\right.
\eeq
acting on the two symbols 0 and 1, taking the symbol 0 as a seed.
More explicitly, the recursion is initiated with a sequence of length one
having only one symbol 0.
Then, while moving along the sequence from left to right, one substitutes 0
by 01 and 1 by 0 until the last digit is reached.
One then repeats the same procedure {\it ad infinitum},
thus generating the infinite Fibonacci sequence.
The first steps of the recursion yield the words 0, 01, 010, 01001, 01001010, $\dots$
The lengths of these finite sequences are given by the consecutive Fibonacci numbers:
1, 2, 3, 5, 8, 13, $\dots$
Alternatively, the $n$-th symbol of the Fibonacci sequence
is given by the explicit formula
\beq
V_n=\left\{\begin{array}{lll}
1\hfill&\hbox{if}&0<\Frac(n\tau)<\tau^{-2},\\
0\hfill&\hbox{if}&\tau^{-2}<\Frac(n\tau)<1,
\end{array} \right.
\label{fibvn}
\eeq
where $\Frac(x)=x-\Int(x)$ denotes the fractional part of $x$, and
\beq
\tau=\frac{\sqrt{5}+1}{2}=1.618033
\eeq
is the golden mean, such that $\tau^2=\tau+1$.
The Fibonacci sequence is therefore both self-similar and quasiperiodic.
It has become popular in the physics literature because it is a
one-dimensional analogue of quasicrystals, discovered in 1984~\cite{ref:qc}.
The density of zeros, \ie, rungs, and ones, \ie, impurities, along the infinite
Fibonacci ladder read $p=\tau^{-1}=\tau-1=0.618033$
and $q=\tau^{-2}=2-\tau=0.381966$, respectively.

\begin{figure}[!ht]
\begin{center}
\includegraphics[angle=-90,width=.5\linewidth]{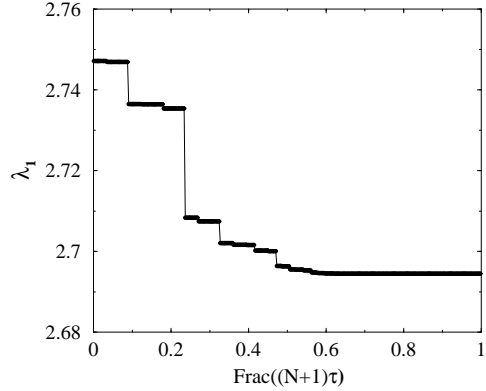}
\end{center}
\caption{\label{fib1}
Plot of the largest eigenvalue $\lambda_1$ of the Fibonacci ladder of size $N$
against the phase $\Frac((N+1)\tau)$.}
\end{figure}

We have considered finite Fibonacci ladders of variable length $N$ (not necessarily even),
with periodic boundary conditions,
where the positions of the rungs is dictated
by the impurity potentials $V_n$ from Eq.~(\ref{fibvn}) for $n=1,\dots,N$.
A numerical diagonalization of the corresponding adjacency matrices
leads to the following observations.
The largest eigenvalue $\lambda_1$ keeps oscillating as a function
of the ladder size $N$
between the asymptotic bounds $\lambda_\min\approx2.6945$ and $\lambda_\max\approx2.7472$.
These oscillations appear as regular if $\lambda_1$
is plotted against the phase $\Frac((N+1)\tau)$ (see Figure~\ref{fib1}).
The modulation of the largest eigenvalue $\lambda_1$ as a function of $N$
therefore follows the quasiperiodicity of the underlying sequence.

\begin{figure}[!ht]
\begin{center}
\includegraphics[angle=-90,width=.39\linewidth]{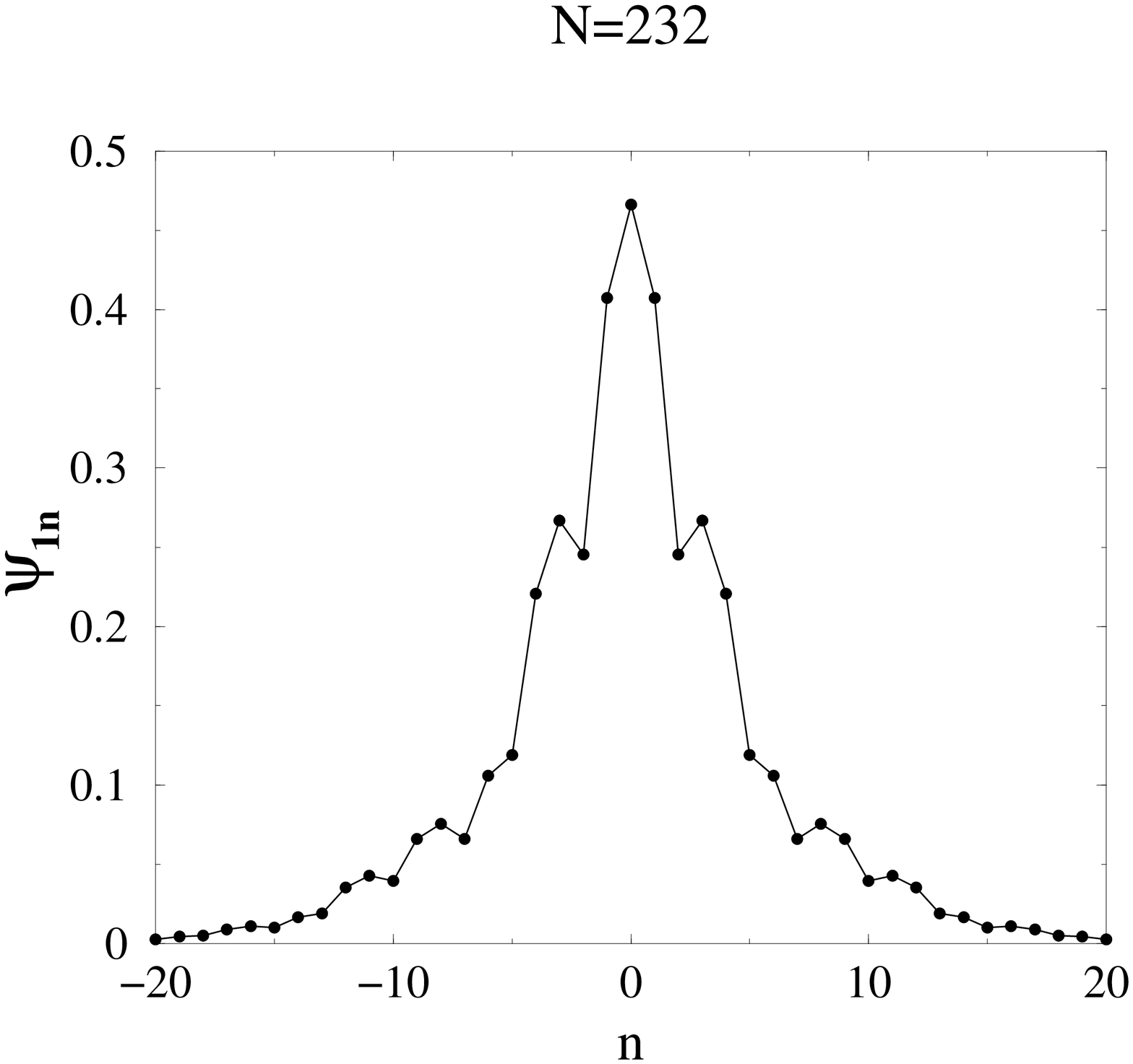}
\includegraphics[angle=-90,width=.4\linewidth]{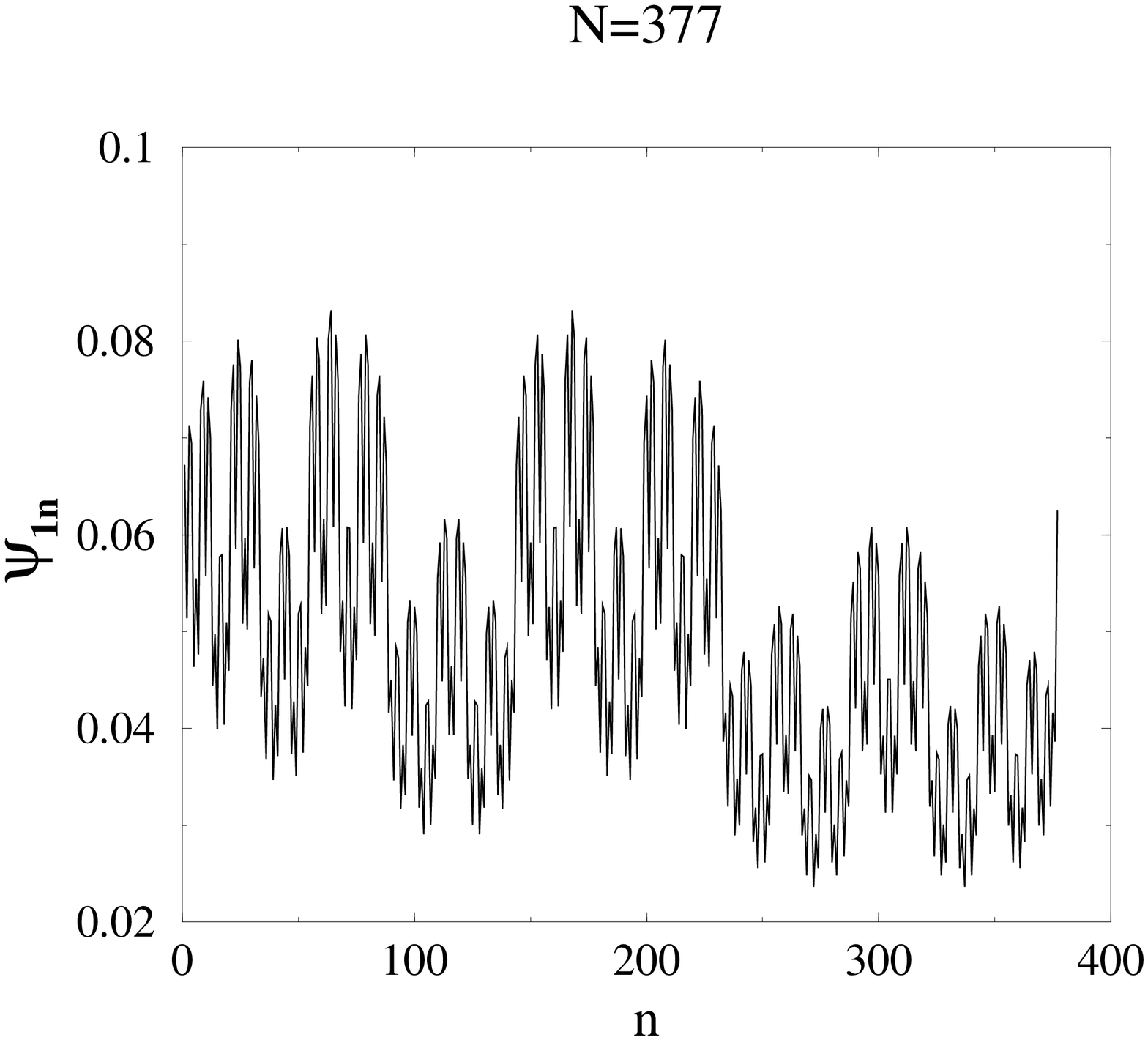}
\end{center}
\caption{\label{fibeigen}
Plot of the ground-state wavefunction $\psi_{1n}$ in two typical examples
of finite Fibonacci ladders.
Left: $N=232$, $\lambda_1=2.747147$.
Right: $N=377$, $\lambda_1=2.694529$.}
\end{figure}

The ground state (Perron-Frobenius eigenvector) $\psi_{1n}$
also exhibits irregular features.
Its appearance varies from localized to extended as a function of the system size $N$.
Figure~\ref{fibeigen} shows one typical example of each kind.
As a general rule, the ground state looks pretty localized
when $\lambda_1$ is close to $\lambda_\max$,
\ie, when $\Frac((N+1)\tau)$ is close to zero.
This is illustrated by the left panel of the figure,
where $N=232$ and $\lambda_1=2.747147\approx\lambda_\max$:
the ground state appears as a symmetric impurity state localized at the boundary,
\ie, around $n=0$.
On the other hand, the ground state looks pretty extended when $\lambda_1$ is close to $\lambda_\min$,
\ie, when $\Frac((N+1)\tau)$ is far from zero.
This is illustrated by the right panel,
where $N=377$ and $\lambda_1=2.694529\approx\lambda_\min$:
the ground state exhibits quite some structure,
but it extends more or less uniformly over the whole ladder.
Let us however recall that the tight-binding Hamiltonian~(\ref{tb1})
on the Fibonacci chain
is known from a rigorous viewpoint to have a purely singular continuous spectral measure,
so that its eigenstates are neither extended nor localized
(see~\cite{ref:sc} for a review).
Generic eigenstates are observed to be multifractal.

\section{Two-dimensional examples}
\label{sec7}

In this last section, we pursue our study of MERW on extended structures
by considering a few two-dimensional situations of interest.

\subsection{Diffusion on the dual $(4,8^2)$ lattice}

\begin{figure}[!ht]
\begin{center}
\includegraphics[angle=90,width=.5\linewidth]{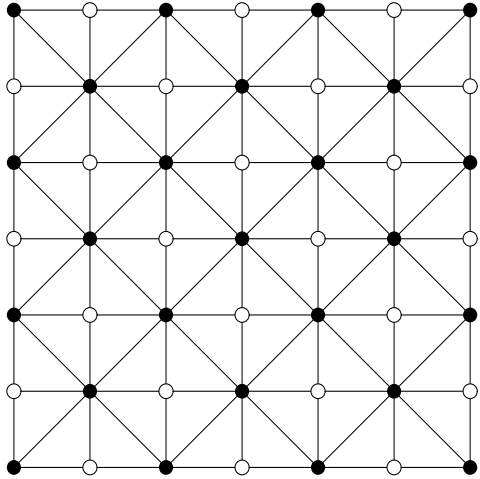}
\caption{\label{lat}
A finite sample of the dual $(4,8^2)$ lattice.}
\end{center}
\end{figure}

We start by an investigation of the transport properties associated with GRW
and MERW
on infinite periodic lattices.
The simplest non-trivial two-dimensional example of a non-bipartite periodic
lattice
is shown in Figure~\ref{lat}.
This lattice is dual to the $(4,8^2)$ Archimedean
lattice~\cite{ref:kepler,ref:gs}.
Nodes denoted by ``$\n$'' with degree $k_\n=8$ and nodes denoted by ``$\b$'' with degree $k_\b=4$ have equal
densities.
It can be checked that $\lambda_1=2(\sqrt{5}+1)$, whereas
$\psi_{1\n}:\psi_{1\b}=(\sqrt{5}+1)/2$.
The effective degrees of the infinite lattice thus~read
\beq
K_1=6,\qquad K_2=2^{8/3}=6.349604,\qquad K_3=2(\sqrt{5}+1)=6.472135.
\eeq

Both GRW and MERW can be viewed as two special cases of the one-parameter
family
of discrete-time random walks defined by the following hopping probabilities onto
neighboring nodes:
\beq
P_{\n\n}=\frac{1+\alpha}{8},\qquad
P_{\b\n}=\frac{1-\alpha}{8},\qquad
P_{\n\b}=\frac{1}{4},
\eeq
where the parameter $\alpha$ is in the range $-1\le\alpha\le1$.
GRW and MERW can be shown to correspond to $\alpha=0$ and
$\alpha=\sqrt{5}-2 = 0.236068$, respectively.

In order to investigate transport properties,
it is advantageous to rewrite the master equation (\ref{master}) in Fourier space.
Denoting by $x$ and $y$ the coordinates on the lattice, we set
\beq
p_\n(x,y,\tau)=\int\frac{\d u\d v}{(2\pi)^2}\,\w p_\n(u,v,\tau)\,\e^{-\i(xu+yv)},
\eeq
and similarly for $p_\b(x,y,\tau)$,
where $u$ and $v$ are the components of the wavevector,
and the integral runs over the first Brillouin zone ($-\pi<u, v<\pi$).
From Eq.~(\ref{master}) we obtain
\beq
\pmatrix{\w p_\n(u,v,\tau+1)\cr\w p_\b(u,v,\tau+1)}
=\Omega(u,v)\pmatrix{\w p_\n(u,v,\tau)\cr\w p_\b(u,v,\tau)},
\eeq
with
\beq
\Omega(u,v)
=\pmatrix{4P_{\n\n}\cos u\cos v&2P_{\n\b}(\cos u+\cos v)\cr 2P_{\b\n}(\cos
u+\cos v)&0}.
\eeq
In the long-wavelength limit, \ie, for small $u,v$,
the largest eigenvalue of this dynamical matrix departs from unity according to
\beq
\omega(u,v)\approx1-\frac{u^2+v^2}{3-\alpha}+\cdots
\eeq
This behavior demonstrates that isotropic diffusion is recovered for all values of
$\alpha$.
The corresponding diffusion constant,
\beq
D(\alpha)=\frac{1}{3-\alpha},
\eeq
increases as a function of $\alpha$ from $1/4$ to $1/2$.
Its values for GRW and MERW read
\beq
D_\GRW=\frac{1}{3}=0.333333,\qquad
D_\MERW=\frac{5+\sqrt{5}}{20}=0.361803.
\eeq
This example shows that MERW can lead to a higher diffusion constant,
\ie, better transport properties, than GRW on some periodic lattices.

\subsection{Designed patterns}

In order to investigate the effect of impurities on MERW in higher dimensions,
before going to the disordered situation of a diluted lattice,
we find it interesting to first look at MERW on designed patterns
consisting of simple geometrical shapes drawn on purpose.

\begin{figure}[!ht]
\begin{center}
\includegraphics[width=6cm]{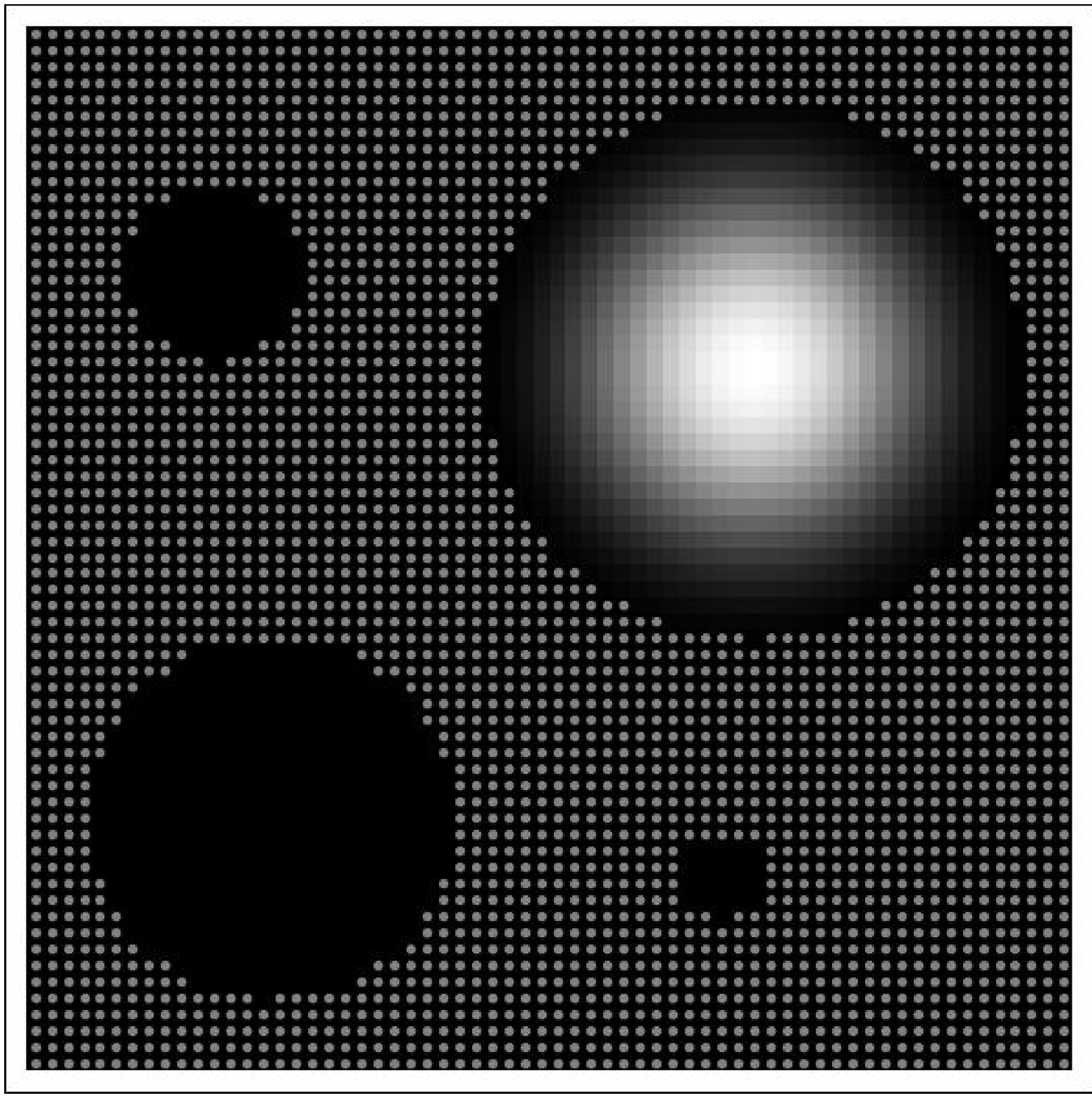}
\includegraphics[width=6cm]{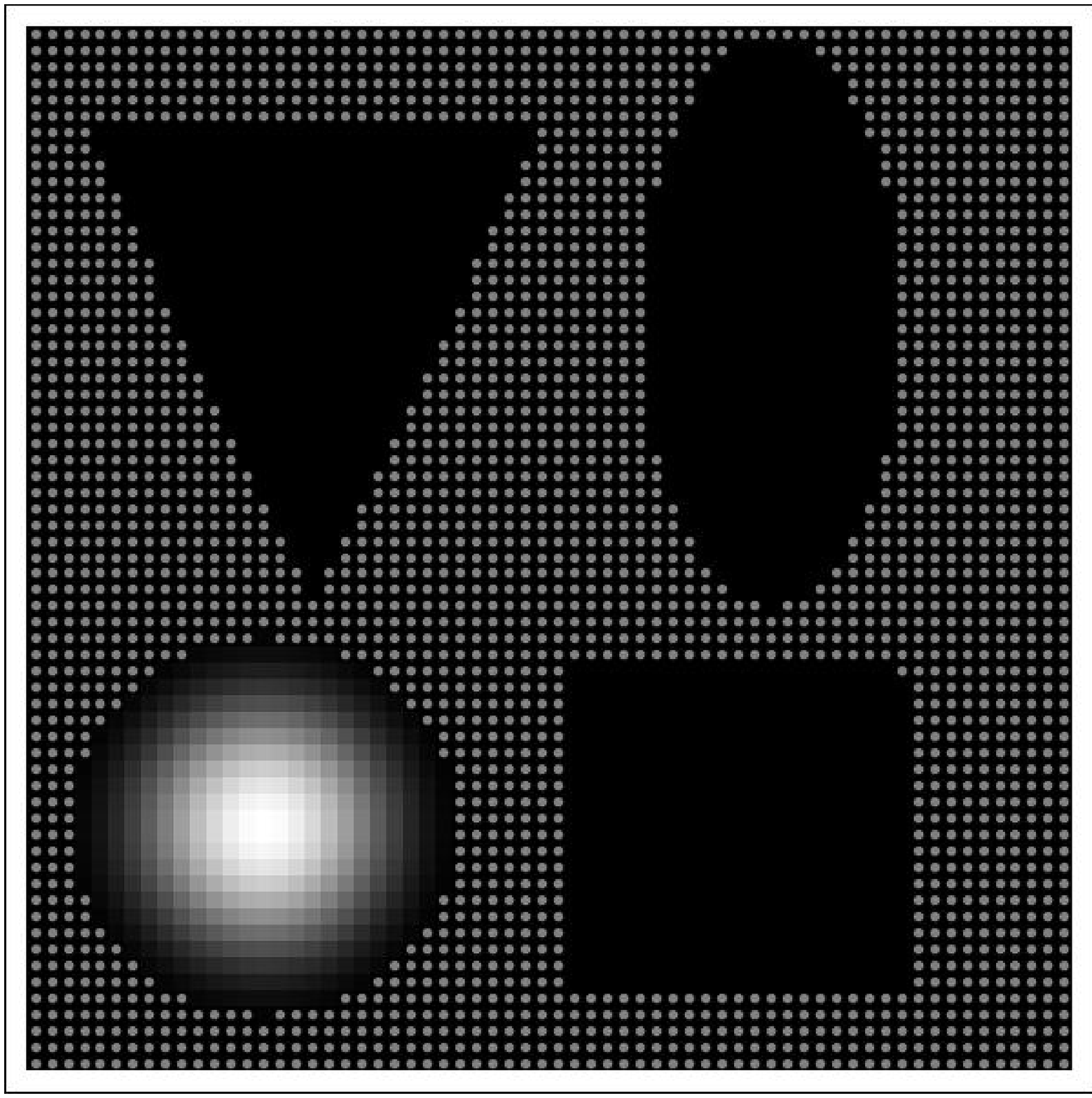}
\end{center}
\caption{\label{fig5_5}
Two examples of designed patterns
drawn on a square lattice with periodic boundary conditions.
Textured regions have a constant unit repulsive potential.
The stationary distribution of MERW is shown as levels of gray
(white: highest probability, black: zero probability).}
\end{figure}

Figure~\ref{fig5_5} shows two examples of patterns
drawn on a square lattice with periodic boundary conditions.
The fully connected square lattice is left untouched in some regions,
so as to have the maximal degree 4, and hence no repulsive potential,
whereas every second horizontal bond is removed in some other regions,
so as to have degree 3, and hence a constant repulsive potential of unit magnitude.
Consider first the left panel of Figure~\ref{fig5_5}.
The fully connected regions are four circles of various sizes.
The stationary distribution $\pi_a$,
evaluated by numerically diagonalizing the adjacency matrix,
is shown as levels of gray.
It is clearly visible that the stationary MERW distribution gets
localized in the largest circle.
Another numerical experiment is shown in the right panel of Figure~\ref{fig5_5}.
This time we have four different shapes, all of them having the same area.
The stationary distribution is observed to localize in the circular region.
To sum up, given a set of domains without defects (where nodes have the highest
degree),
a particle performing MERW tends to spend most of its time,
and gets eventually localized, in the largest and most circular of these~domains.

\subsection{The diluted square lattice: stationary state}

We now turn to the case of the diluted square lattice,
where bonds are removed at random with a (small) probability $q$.

The Lifshitz argument presented in Section~\ref{sec:lif}
generalizes to higher dimensions~\cite{ref:mplif,ref:trap}.
The ground state is expected to be localized
in the largest {\it Lifshitz region}, \ie, nearly circular region free of defects.
The quantitative analysis of the phenomenon goes as follows.
The radius $R_1$ of the largest Lifshitz region
can be estimated by means of extreme-value statistics.
On a finite system of size $L\times L$,
the number of nearly circular regions of radius~$R$ with no defect
is of order $L^2 p^{2\pi R^2}$, as there are two links per node.
The criterion that this number becomes of order unity yields
\beq
R_1\approx\left(\frac{\ln L}{\pi\abs{\ln p}}\right)^{1/2}.
\label{lifr}
\eeq
Now, using a continuum description, the ground state in the disk of radius
$R_1$ with Dirichlet boundary conditions
is given by $\psi(r)\sim J_0(jr/R_1)$, where~$r$ is the distance from the center,
and $j=2.404825\dots$ is the first zero of the Bessel function $J_0$.
We thus obtain
\beq
E_1\approx\left(\frac{j}{R_1}\right)^2\approx\frac{\pi j^2\abs{\ln p}}{\ln L}.
\eeq
The corresponding thermodynamical statement, generalizing (\ref{lifrho}), reads
\beq
\rho(E)\sim\exp\left(-\frac{2\pi j^2\abs{\ln p}}{E}\right).
\eeq

\begin{figure}[!ht]
\begin{center}
\includegraphics[width=6cm]{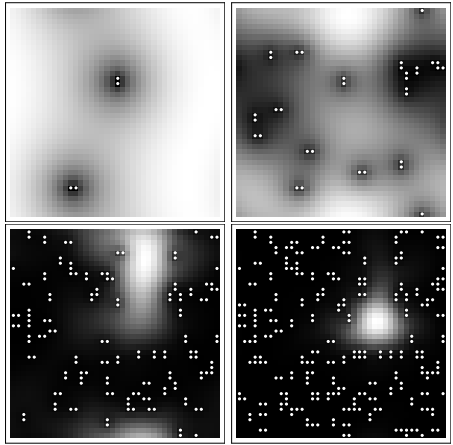}
\end{center}
\caption{\label{fig7}
Density plots of the stationary distribution for MERW on a $40\times 40$
square lattice with periodic boundary conditions,
for the concentrations $q=0.001,0.01,0.05,0.1$ of removed links.
The endpoints of removed links are shown as bright dots.
The stationary probability is shown as levels of gray.}
\end{figure}

This picture is corroborated by the data shown in Figure~\ref{fig7}.
The stationary probability, shown as levels of gray,
is observed to be localized in the largest nearly circular regions free
of defects. Their sizes are in rough agreement with the estimate (\ref{lifr}),
yielding, respectively, $R_1=34.3$, 10.8, 4.78 and 3.34 for $q=0.001$, 0.01,
0.05 and 0.1.

In higher dimension ($d\ge2$), skipping lattice-dependent multiplicative constants,
the above estimates become
\beq
-\ln\rho(E)\sim\frac{\abs{\ln p}}{E^{d/2}},
\eeq
and
\beq
R_1\sim\left(\frac{\ln L}{\abs{\ln p}}\right)^{1/d},
\qquad
E_1\sim\left(\frac{\abs{\ln p}}{\ln L}\right)^{2/d},
\eeq
where $L$ is the linear size of the sample.
Hence the stationary distribution of MERW
on a sufficiently large $d$-dimensional lattice in the presence of any finite concentration $q$ of disorder
is localized in the largest nearly spherical Lifshitz region
whose volume grows as $\ln L$.

Let us close up with a word on the weak-disorder crossover.
When the amount of disorder gets smaller and smaller ($q\ll1$),
the volume of the Lifshitz sphere diverges as $\Omega_1\sim R_1^d\sim(\ln
L)/q$.
When this estimate is of the order of the volume $L^d$ of the sample,
MERW experiences a crossover between
a Lifshitz localized regime (for $L^d\gg\Omega_1$, \ie, $qL^d\gg\ln L$)
and an extended regime (for $L^d\ll\Omega_1$, \ie, $qL^d\ll\ln L$).
The number $qL^d$ of impurities needed to drive the crossover is therefore very modest,
as it also grows as $\ln L$.

\subsection{The diluted square lattice: dynamics}

So far, we have seen that the stationary distributions for GRW and MERW
are qualitatively different in the presence of disorder, such as a weak dilution.
A particle performing MERW will eventually end up
in the largest Lifshitz sphere, \ie, the largest nearly spherical region free
of defects,
where the stationary distribution is localized.
Inside this region, MERW will look much like usual random walk,
\ie, like Brownian motion on large space and time scales.

In the transient regime, \ie, before the particle finds its stationary state,
it is expected to stay for a while in some Lifshitz region,
smaller than the optimal one but nearer to its starting point,
where it will spend some time, before it ``learns'' that there is a better,
albeit more distant, region elsewhere in the system, and so on.
The diffusion process will thus explore a sequence of consecutive metastable states,
depending on the initial point, before finally reaching the true ground state.
A similar picture should hold for a diffusive particle in the presence
of a random distribution of absorbing traps~\cite{ref:trap},
if the observation is conditioned on the survival of the particle.
We can therefore expect the dynamics to exhibit two different time scales,
just as in many glassy systems~(see \eg~\cite{ref:glassy}):
a fast {\it (beta)} relaxation within each metastable Lifshitz region
(where the entropy can be maximized locally),
and a slow {\it (alpha)} relaxation
corresponding to tunneling between consecutive Lifshitz regions,
until the optimal region which carries the true ground state is reached,
so that the entropy production rate has attained its maximal value.

\begin{figure}[!ht]
\begin{center}
\includegraphics[width=3cm]{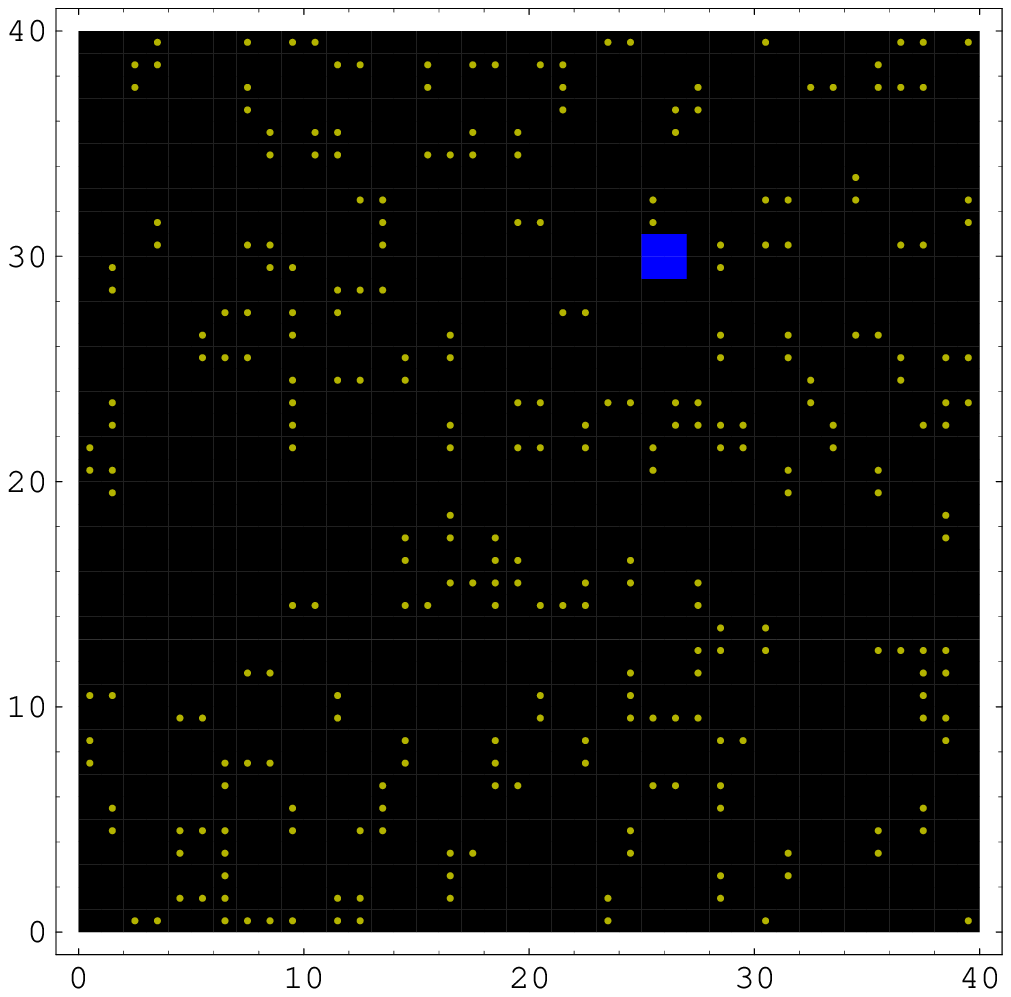}
\includegraphics[width=3cm]{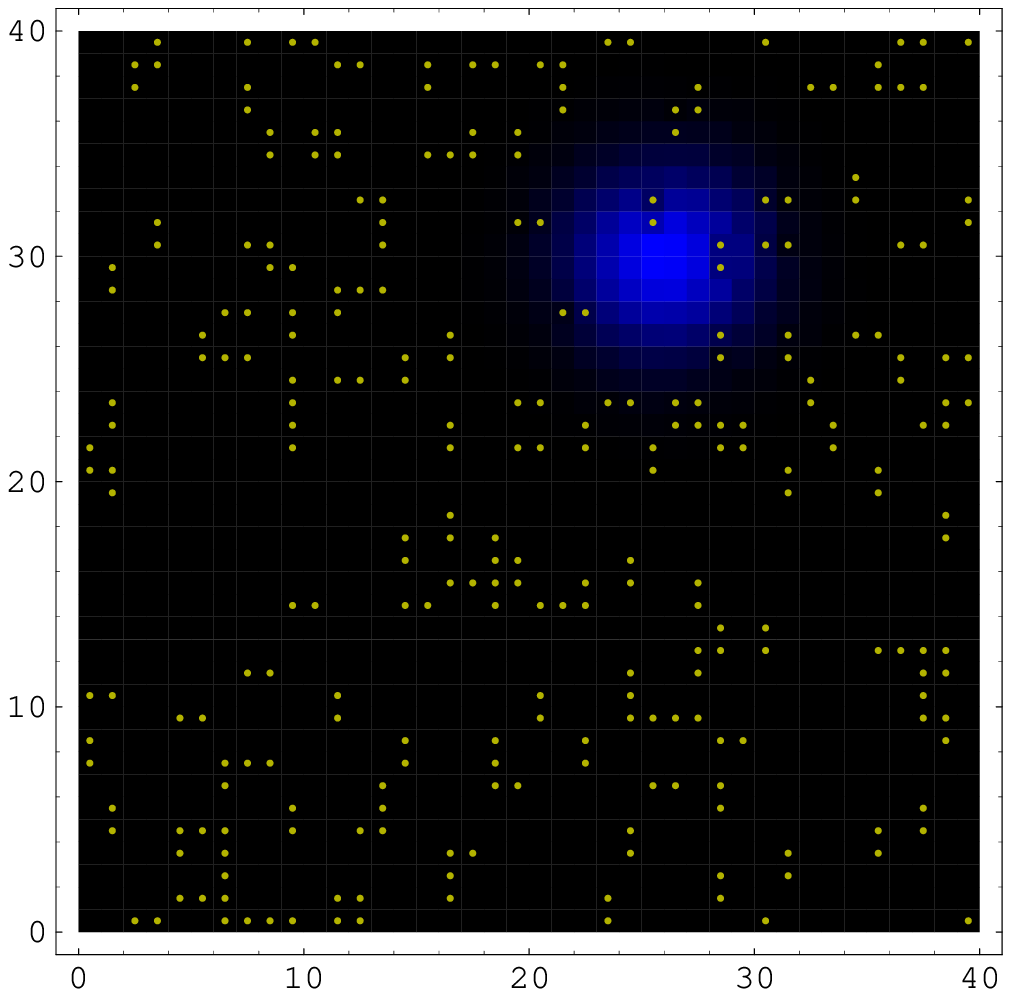}
\includegraphics[width=3cm]{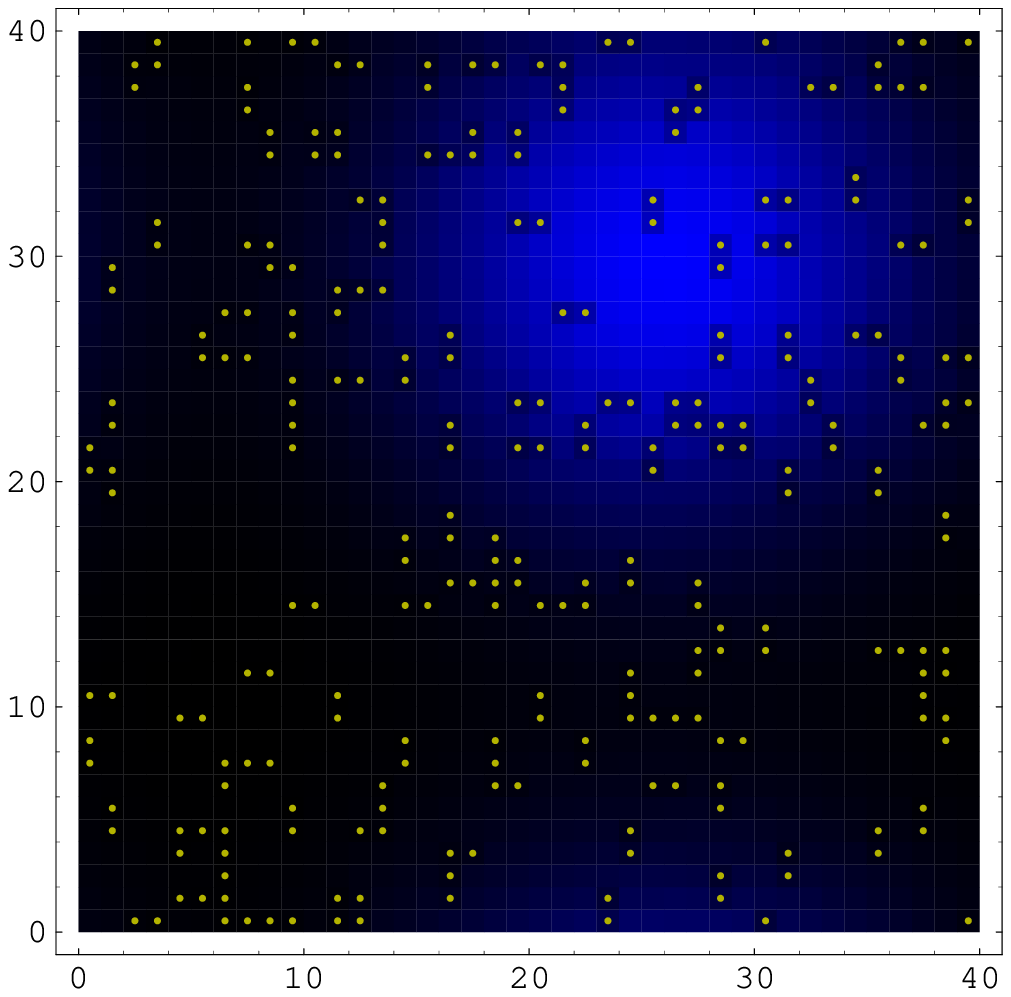}
\includegraphics[width=3cm]{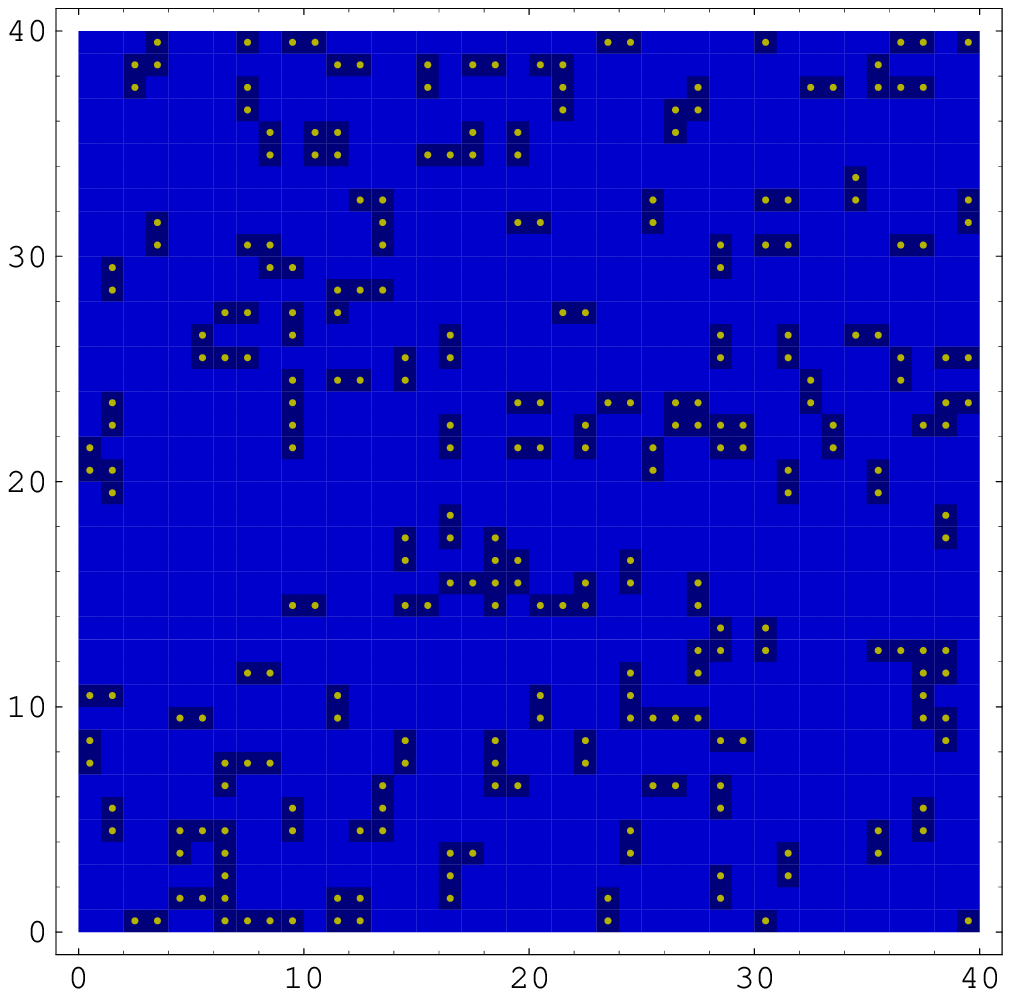}
\includegraphics[width=3cm]{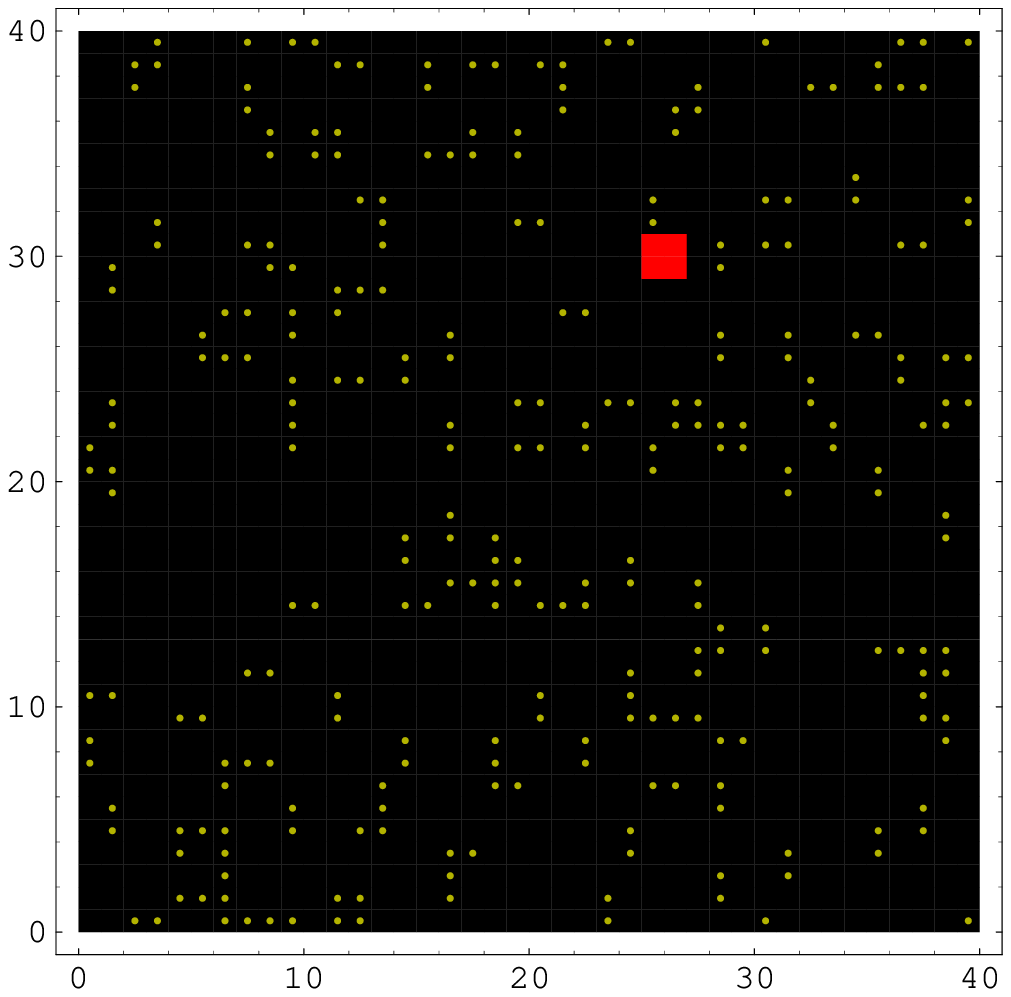}
\includegraphics[width=3cm]{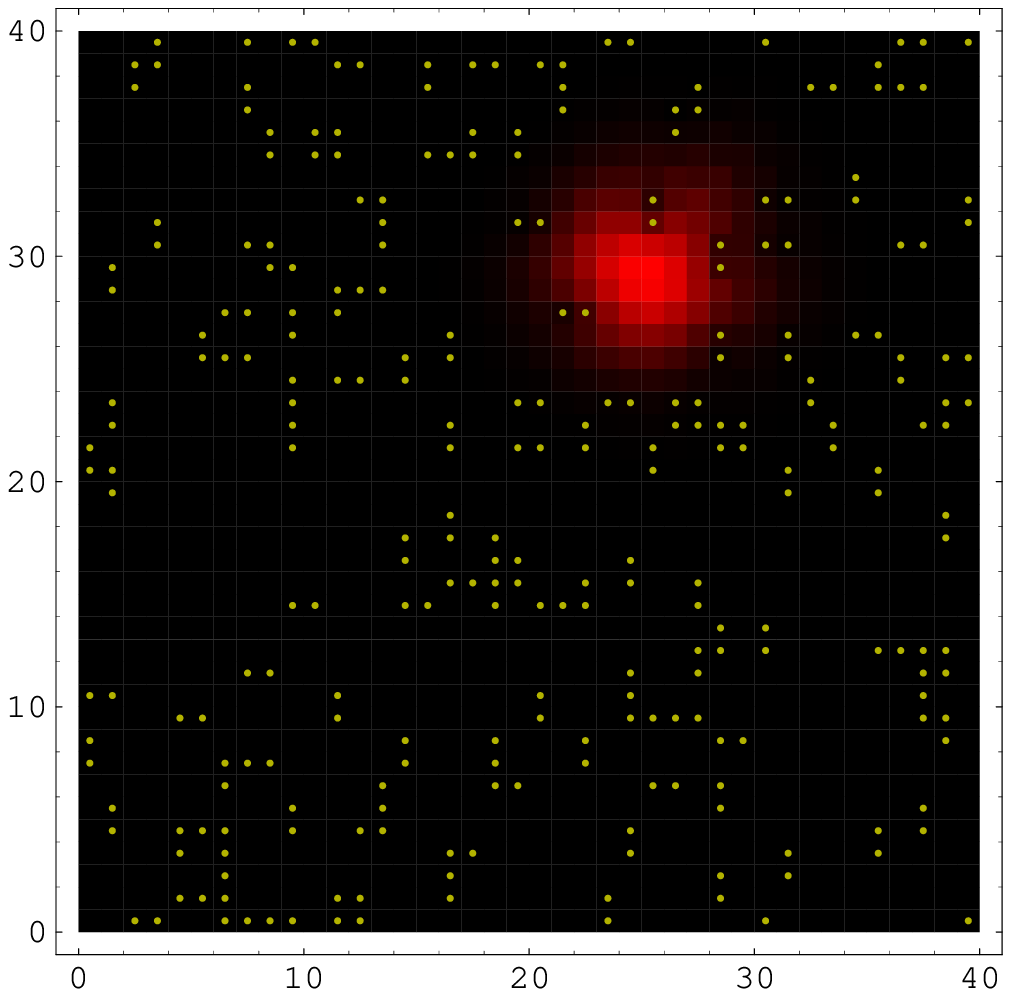}
\includegraphics[width=3cm]{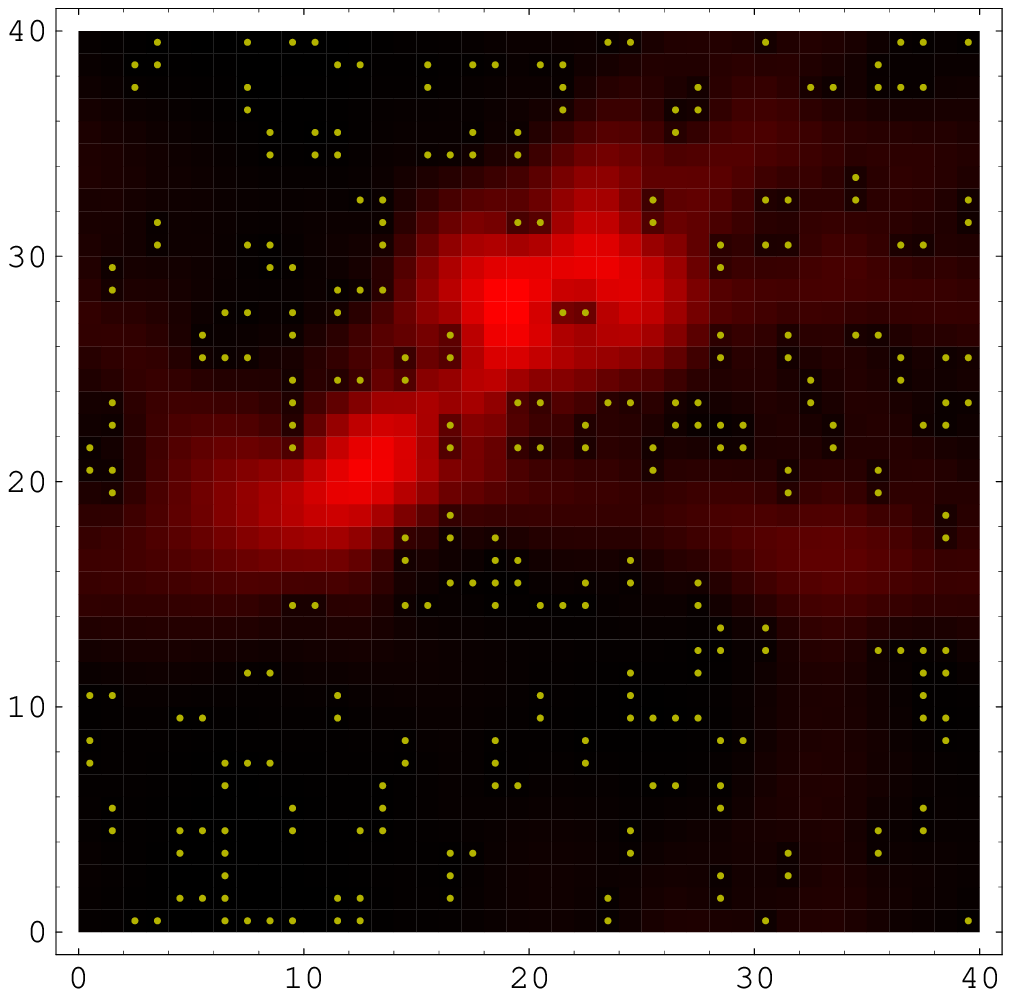}
\includegraphics[width=3cm]{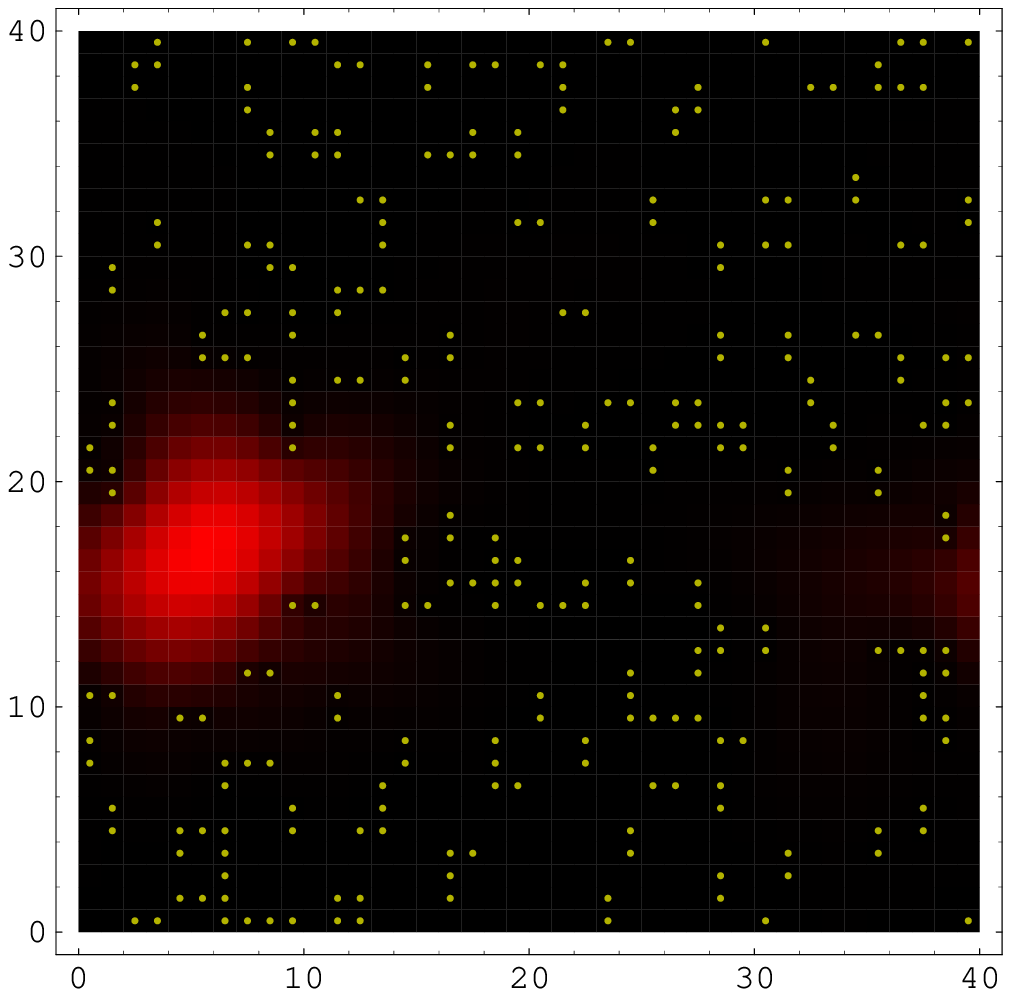}
\end{center}
\caption{\label{fig8}
(Color online).
Evolution of GRW (upper panels) and MERW (lower panels) on a $40\times40$
square lattice with a concentration $q=0.1$ of removed links.
Left to right: distribution after 1, 16, 128 time steps and in the stationary
state.}
\end{figure}

This difference between the dynamical evolution of GRW and MERW
is illustrated in Figure~\ref{fig8}.
A particle performing GRW (upper panels) and MERW (lower panels)
starts at the same site
of a lattice with the same randomly positioned defects. In the course of
evolution,
the generic random walker visits every site with a probability proportional to
its degree.
After a sufficiently long time, the probability distribution
spreads more or less uniformly over the lattice.
The stationary distribution is locally modulated by the presence of impurities,
but it remains globally extended, just as in the case of a lattice without defects.
The situation for the maximal entropy random walker is quite
different. The particle indeed visits a sequence of larger and larger,
albeit more and more distant regions free of defects,
until it eventually reaches its stationary state.

\subsubsection*{Acknowledgments}

It is a pleasure to thank St\'ephane Nonnenmacher for having made us aware of
the concept of Shannon-Parry measures known in ergodic theory.
BW acknowledges partial support by the EPSRC grant EP/030173
and ZB by the Polish Ministry of Science Grant No. N N202 229137 (2009-2012).

\end{document}